\documentclass{mn2e}
\usepackage{epsf,times}
\newcommand{\etal}{{et al}\/.}

\begin{document}
\title[{\rm Chandra} and {\rm XMM} observations of 3C\,465]{A {\it
    Chandra} and {\it XMM-Newton} study of the wide-angle tail radio
    galaxy 3C\,465}
\author[M.J.~Hardcastle \etal]{M.J.\ Hardcastle$^{1,2}$, I. Sakelliou$^{3,4}$ and D.M.\ Worrall$^2$
\\
$^1$ School of Physics, Astronomy and Mathematics, University of
Hertfordshire, College Lane, Hatfield, Hertfordshire AL10 9AB\\
$^2$ Department of Physics, University of Bristol, Tyndall Avenue,
Bristol BS8 1TL\\
$^3$ Max-Planck-Institut f\"ur Astronomie, K\"onigstuhl 17, D-69117
    Heidelberg, Germany\\
$^4$ School of Physics and Astronomy, University of Birmingham, Edgbaston, Birmingham B15 2TT\\
}
\maketitle
\begin{abstract}
We have observed the prototypical wide-angle tail radio galaxy 3C\,465
with {\it Chandra} and {\it XMM-Newton}. X-ray emission is detected
from the active nucleus and the inner radio jet, as well as a
small-scale, cool component of thermal emission, a number of the
individual galaxies of the host cluster (Abell 2634), and the hotter
thermal emission from the cluster itself. The X-ray detection of the
jet allows us to argue that synchrotron emission may be an important
mechanism in other well-collimated, fast jets, including those of
classical double radio sources. The bases of the radio plumes are not
detected in the X-ray, which supports the model in which these plumes
are physically different from the twin jets of lower-power radio
galaxies. The plumes are in fact spatially coincident with deficits of
X-ray emission on large scales, which argues that they contain little
thermal material at the cluster temperature, although the minimum
pressures throughout the source are lower than the external pressures
estimated from the observed thermal emission. Our observations confirm
both spatially and spectrally that a component of dense, cool gas with
a short cooling time is associated with the central galaxy. However,
there is no evidence for the kind of discontinuity in external
properties that would be required in many models of the jet-plume
transition in WATs. Although the WAT jet-plume transition appears
likely to be related to the interface between this central cool
component and the hotter intra-cluster medium, the mechanism for WAT
formation remains unclear. We revisit the question of the bending of
WAT plumes, and show that the plumes can be bent by plausible bulk
motions of the intra-cluster medium, or by motion of the host galaxy
with respect to the cluster, as long as the plumes are light.
\end{abstract}

\section{Introduction}

Wide-angle tail radio galaxies (WATs) present a number of interesting
problems for our understanding of the dynamics of extragalactic radio
sources and in particular their interaction with the external medium.
Here we define the class as being sources that are associated with a
cluster dominant galaxy, at or near the cluster centre, and that are
characterized in the radio by twin, well-collimated jets that abruptly
flare into long, often sharply bent plumes or tails. (Throughout this
paper we use the term `jet' to refer to the inner, well-collimated
feature, and `plume' to refer to the outer parts of the source.) The
earliest problem associated with WATs was identified by Burns (1981)
and Eilek \etal\ (1984, hereafter E84): given that WAT host galaxies
are cluster-centre objects and should be close to the bottom of the
cluster potential well (and so not moving at high speed with respect
to the external medium) what gives rise to the often sharp bends in
the plumes? This problem is in fact not unique to sources that are
WATs by our definition; not all cluster-centre sources with bent
plumes exhibit the large-scale, narrow, well-collimated jets of a WAT.
A problem that {\it is} unique to WATs (also identified by E84) is
how, and why, these jets make the transition into broad plumes where
they do: what is it about the external environment that sets the
location for the jet-plume transition? Or, since these objects are in
general radio-luminous enough to be classical double (FRII) sources
and their jets are very similar to those of FRIIs, we can ask a
related question (e.g. Hardcastle 1998): what is it about the external
environment that causes these sources to be WATs and not FRIIs?

Radio, optical and X-ray observations have allowed some progress to be
made in answering these questions. Large-scale bulk motions in the
intra-cluster medium (ICM) are expected if the host cluster is not
relaxed, and the ram pressure from these can bend the plumes: {\it
Einstein} and {\it ROSAT} observations have shown that there is indeed
X-ray substructure in the host clusters of WATs (e.g. Burns \etal\
1994; G\'omez \etal\ 1997a) and this is backed up by optical
spectroscopy that shows there is substantial substructure in the
galaxy velocity distributions of a few well-studied clusters hosting
WATs (Pinkney \etal\ 1993; G\'omez \etal\ 1997b). Substructure in the
host clusters will also give rise to non-radial buoyancy forces that
can play a part in bending the plumes (e.g. Sakelliou, Merrifield \&
McHardy 1996). Observational input to the questions related to the
jet-plume transition has been more limited, but Hardcastle (1998) and
Hardcastle \& Sakelliou (2004) have shown that the jets often
terminate in compact structures similar to the hotspots of FRII radio
galaxies, supporting a picture (Hardcastle 1999) in which the current
jet-plume transition, where it occurs at a single point, is a result
of the interaction between the jet and the edge of the previously
existing plume. Hardcastle \& Sakelliou also provided the first direct
evidence that the external environment is responsible for the unique
structure of WATs, by demonstrating an inverse correlation between the
jet termination distance and the temperature of the host cluster.
However, the physics that governs the location of the base of the
plume is still unclear. Modellers have concentrated on the idea that
it is associated with some localized feature of the external medium,
such as propagation across an IGM/ICM interface or a shock front (e.g.
Loken \etal\ 1995; Hooda \& Wiita 1996) or interaction with discrete
clumps in the external medium (e.g. Higgins, O'Brien, \& Dunlop 1999).
X-ray observations with the previous generation of observatories had
insufficient resolution and sensitivity to test these models.

However, {\it Chandra} and {\it XMM} have now been used to observe a
number of WAT host clusters, allowing the observational picture in the
X-ray to be updated. Jetha \etal\ (2005) present observations of two
WAT host clusters with {\it Chandra}, showing that both contain a cool
central gas component and that the jet enters the plume close to the
transition between this and the hotter gas in the rest of the host
cluster. While this has some similarities to the picture in which a
jet-plume transition is a result of a propagation across some
structure in the IGM/ICM interface, the detailed {\it Chandra}
observations show that there are no discontinuities in the physical
conditions in the clusters (as would be expected if the jet-plume
transition was a result of propagation across a discrete feature in
the external medium, such as a shock or cold front) and no discrete
X-ray emitting features associated with the plume bases. An
association between the jet-plume transition and the cold central gas
seems likely as a result of this work, but the mechanism for WAT formation
is still not clear.

The observations of Jetha \etal\ were also not sensitive enough to
study the X-ray properties of the radio components themselves. In WATs
the jets, like those in FRIIs, are thought to remain relativistic,
well-collimated flows (`type II' jets) until they undergo rapid
deceleration at a hotspot. In the much commoner twin-jet FRI sources,
the initially type II jet flow undergoes significant bulk deceleration
on scales of a few kpc, slowing from bulk relativistic speeds to
speeds that are trans-sonic or sub-sonic with respect to the external
medium (Bicknell 1994; Laing \& Bridle 2002) and forming a `type I'
jet that may in some cases make a smooth transition into a plume
similar to those of WATs. In all cases that have been observed in
detail this bulk deceleration in FRIs is accompanied by high-energy
particle acceleration that gives rise to X-ray synchrotron emission
(e.g. Hardcastle \etal\ 2001, Worrall \etal\ 2001, Hardcastle \etal\
2002). However, the inner jets of FRI sources, while they should still
be relativistic, are often also strong sources of X-ray emission
(Hardcastle \etal\ 2001, 2003). WAT jets should be similar to the
inner jets of FRIs, and so we might expect to detect synchrotron X-ray
emission from them, particularly on small scales. On the other hand,
WAT plumes should {\it not} be similar to the brighter outer jets of
FRIs in the deceleration region, since the model outlined by
Hardcastle \& Sakelliou (2004) requires the jet deceleration in the
plumes to take place at a single discrete location.

Investigating these issues requires sensitive X-ray observations of
more WAT systems. 3C\,465, the subject of this paper, is the closest
WAT in the northern sky ($z=0.0293$) and the best studied at radio
frequencies (Riley \& Branson 1973; Leahy 1984; E84;
Eilek \& Owen 2002; Hardcastle \& Sakelliou 2004), to the extent that
it is often taken to be the prototype object of its class. Its host
galaxy is the D galaxy NGC 7720, the dominant galaxy of the cluster
Abell 2634. A2634 has been well studied in the X-ray, initially with
{\it Einstein} (Jones \etal\ 1979; E84; Jones \& Forman
1984) and later with the {\it ROSAT} PSPC (Schindler \& Prieto 1997)
and HRI (Sakelliou \& Merrifield 1998, 1999) and with ASCA (e.g.
Fukazawa \etal\ 1998). The velocity properties of the galaxies in the
cluster have also been studied in detail (Pinkney \etal\ 1993,
Scodeggio \etal\ 1995). From these studies it has been possible to
show that A2634 is a non-relaxed cluster, with an X-ray extension
around the central cluster galaxy perpendicular to the jet axis (e.g.
Schindler \& Prieto 1997) although there is little evidence in the
galaxy distributions and velocity data for a recent merger (Scodeggio
\etal\ 1995). On small scales, Sakelliou \& Merrifield (1999) argued
that the compact X-ray core seen by the {\it ROSAT} HRI was extended
on scales of a few arcsec, and that this implied detection of a
small-scale IGM component. Sakelliou \& Merrifield (1998) showed that
the individual galaxies of the host cluster were of low luminosity in
the X-ray, arguing that this implied that hot gas had been stripped
from the galaxies in their motion through the IGM.

Because of this wealth of existing observational data, 3C\,465 is an
excellent target for observations with the new generation of X-ray
observatories aimed at addressing the outstanding questions described
above. In this paper we report on the results of our {\it Chandra} and
{\it XMM-Newton} observations. Throughout the paper we use a
concordance cosmology with $H_0 = 70$ km s$^{-1}$ Mpc$^{-1}$,
$\Omega_{\rm m} = 0.3$ and $\Omega_\Lambda = 0.7$. At the redshift of
3C\,465, 1 arcsec corresponds to 590 pc. Spectral indices $\alpha$ are
the energy indices and are defined in the sense $S_{\nu} \propto
\nu^{-\alpha}$. The photon index $\Gamma$ is $1+\alpha$. All fits to
the X-ray data include the effects of Galactic absorption with a
column density of $4.91 \times 10^{20}$ cm$^{-2}$. J2000 co-ordinates
are used throughout.

\section{Observations and processing}

\subsection{XMM}
\label{xmm-intro}

3C\,465 was observed with the EPIC cameras on {\it XMM}, using the
medium filter, for 11160 s (MOS1) 11162 s (MOS2) and 8725 s (pn) on
2002 Jun 22. The data were filtered using {\sc sas} to include only
the standard flag set, and remove bad columns and rows; we retained
the out of field of view events. High background events were filtered
on the count rate above 10 keV using the same criteria as were used
for the background dataset we used in subsequent analysis (Read \&
Ponman 2003), i.e. an upper count rate in the 10--15 keV energy range
with standard flags and PATTERN=0 of 1 count s$^{-1}$ (pn) and 0.35
count s$^{-1}$ (MOS): this did not significantly affect the MOS cameras, but
flagged out a substantial fraction of the pn data. The effective
exposure times after filtering were 11054 s (MOS1) 11105 s (MOS2) and
6735 s (pn).

After filtering, the data were vignetting-corrected using the {\sc
sas} task {\it evigweight}, and spectra and images were generated
using the {\it evselect} task. We used the techniques described by
Croston \etal\ (2003) to interpolate over chip gaps and bad columns to
make images in the energy range 0.3--7.0 keV for presentation.

Although the source and background datasets had been filtered using
the same criteria, the count rate in the out of field of view regions
for both MOS and pn was significantly higher in our observations than
in the background dataset. This implies a higher contribution from
particles in our data, which (if uncorrected) would bias our extracted
spectra. The reason for this discrepancy appears to be that the
effective spectrum of the particle background, as measured by {\it
XMM}, differs in our data and in the background files (where it is of
course an exposure-weighted average over many observations). We
generated an approximate correction for the particle background excess
in our data by taking the difference between the out of field of view
spectra in our data and those in the Read \& Ponman background files.
A suitably scaled version of this spectrum (taking account of the
weighting applied to the data) was then added to the {\it
evigweight}-corrected background spectra generated for each
observation. This had the effect of stabilizing the
temperature fitted to large-area regions of the extended X-ray source
(Section \ref{cluster}).

While this approach to background correction should provide an
accurate correction for the non-X-ray contribution to the observed
background, it differs from the double-subtraction method of Arnaud
\etal\ (2002) in that it does not attempt to correct for the
difference between the level of the cosmic X-ray background in the
background datasets and that in the observation of 3C\,465. The reason
we do not attempt to do this is that the cluster fills the field of
view, so that it is impossible to define a local blank-sky background
that does not include cluster emission: any attempt to apply the
method of Arnaud \etal\ would tend to over-estimate the cosmic
background excess and consequently cause us to under-estimate the true
normalization of the models fitted to the data. The systematic error
introduced by omitting this double-subtraction correction is small in
the case of 3C\,465 and should predominantly affect low energies, and
we have verified that applying the double-subtraction method to the
spectra we extract in Section \ref{cluster} does not significantly
affect fitted parameters such as {\sc mekal} model temperature, though
of course it does affect the normalization of the fitted models.

\subsection{Chandra}

We observed 3C\,465 with {\it Chandra} on 2004 Aug 31 for 49528 s. The
source was positioned near the standard aim point on the
back-illuminated S3 CCD. We reprocessed the data in the standard way
to apply the latest calibration files (using CIAO 3.1 and CALDB 2.28).
As the data were taken in VFAINT mode, we used `VFAINT cleaning' to
identify and reject some background events. We also removed the
0.5-pixel event position randomization. There were no times of
unusually high background, so we were able to use all of the available data.

We used events in the energy range 0.5--5 keV to construct images
(throughout the paper this energy range is used for {\it Chandra}
imaging).

\subsection{Radio data}

The high-resolution radio images used are taken from Hardcastle \&
Sakelliou (2004), and were made with the NRAO Very Large Array (VLA)
at 8.4 GHz. We also make use of a 1.4-GHz VLA image taken from the
3CRR Atlas of Leahy, Bridle \&
Strom\footnote{http://www.jb.man.ac.uk/atlas/} with a resolution of
5.4 arcsec.

\section{Spatial and spectral analysis}

\subsection{Methods}

{\it Chandra} spectra were extracted and corresponding response and
ancillary response matrices constructed using {\sc ciao}. {\it XMM}
spectral extraction was carried out using SAS, and an on-axis
ancillary response matrix was generated using the task {\it arfgen}
and used for every spectrum. The response matrices used were the
`canned' responses supplied on the {\it XMM} web
site\footnote{ftp://xmm.vilspa.esa.es/pub/ccf/constituents/extras/responses/}.
The use of on-axis ancillary response matrices is made possible by the
vignetting correction applied with {\it evigweight}.

Spectral model fitting was carried out in {\sc xspec}, in the energy
range 0.4--7.0 keV ({\it Chandra}) and 0.3--7.0 keV ({\it XMM}). In
all cases, except where otherwise stated, the extracted spectra were
grouped so that each fitting bin had $>20$ counts after background
subtraction. Errors quoted throughout are the $1\sigma$ value for one
interesting parameter, unless otherwise stated.

\subsection{Overview}

\begin{figure*}
\epsfxsize 18cm
\epsfbox{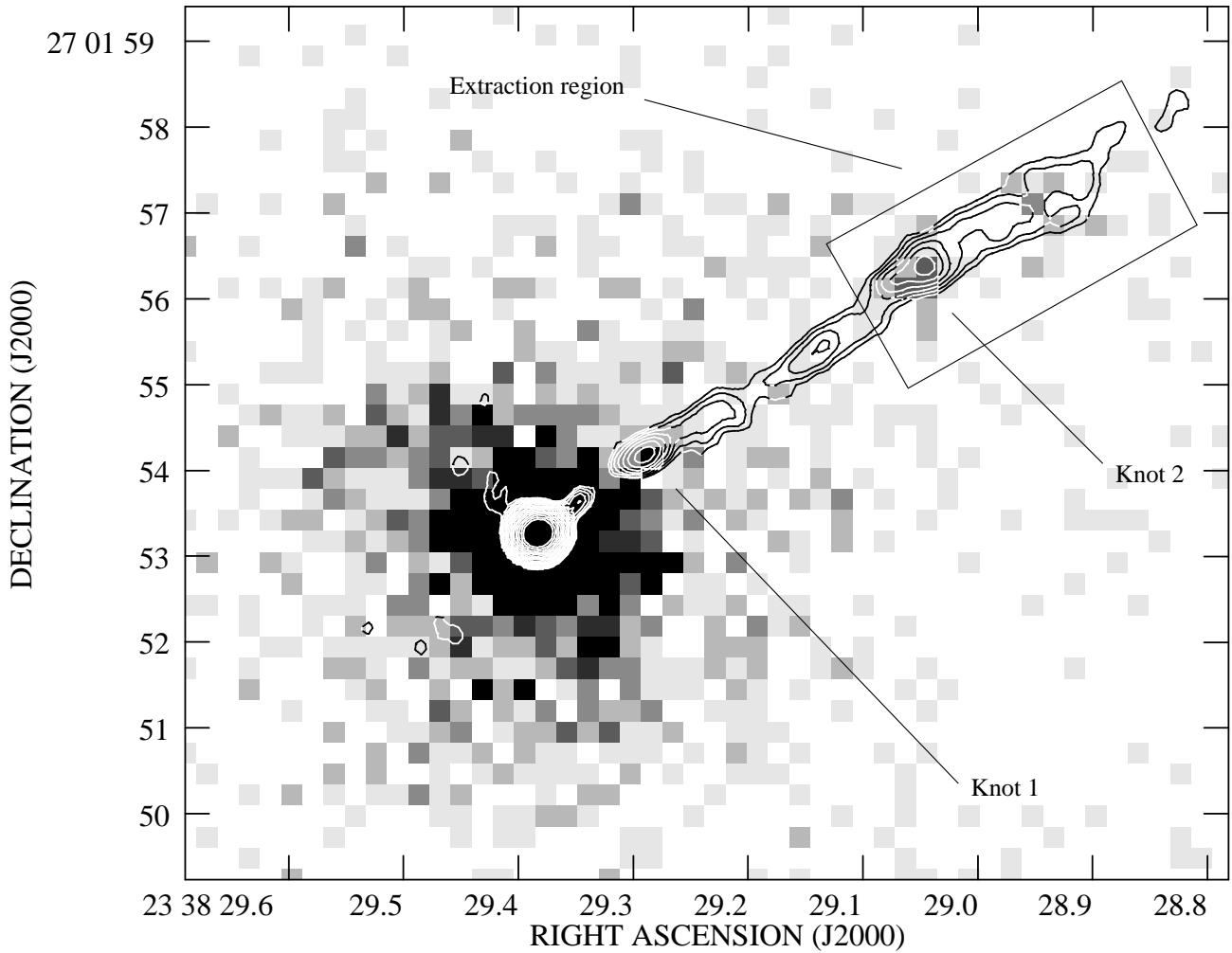}
\caption{Small-scale X-ray emission from the central regions of
  3C\,465. The greyscale shows the 0.5--5.0 keV {\it Chandra} data
  binned in 0.246-arcsec pixels; black is 6 counts pixel$^{-1}$.
  Superposed are contours of a $0.28 \times 0.26$ (major $\times$
  minor axis FWHM) resolution 8.4-GHz VLA map at $0.1 \times (1,
  \sqrt{2}, 2, 2\sqrt{2} \dots)$ mJy beam$^{-1}$. Two knots in the jet
  with possible X-ray detections are marked, and the region used for
  spectral measurements is shown. The X-ray co-ordinates have been
  shifted by around 0.2 arcsec to align the X-ray centroid with the
  radio core: this shift is well within the uncertainties on {\it
  Chandra} astrometric accuracy.}
\label{chandra-image}
\end{figure*}

\begin{figure*}
\epsfxsize 17cm
\epsfbox{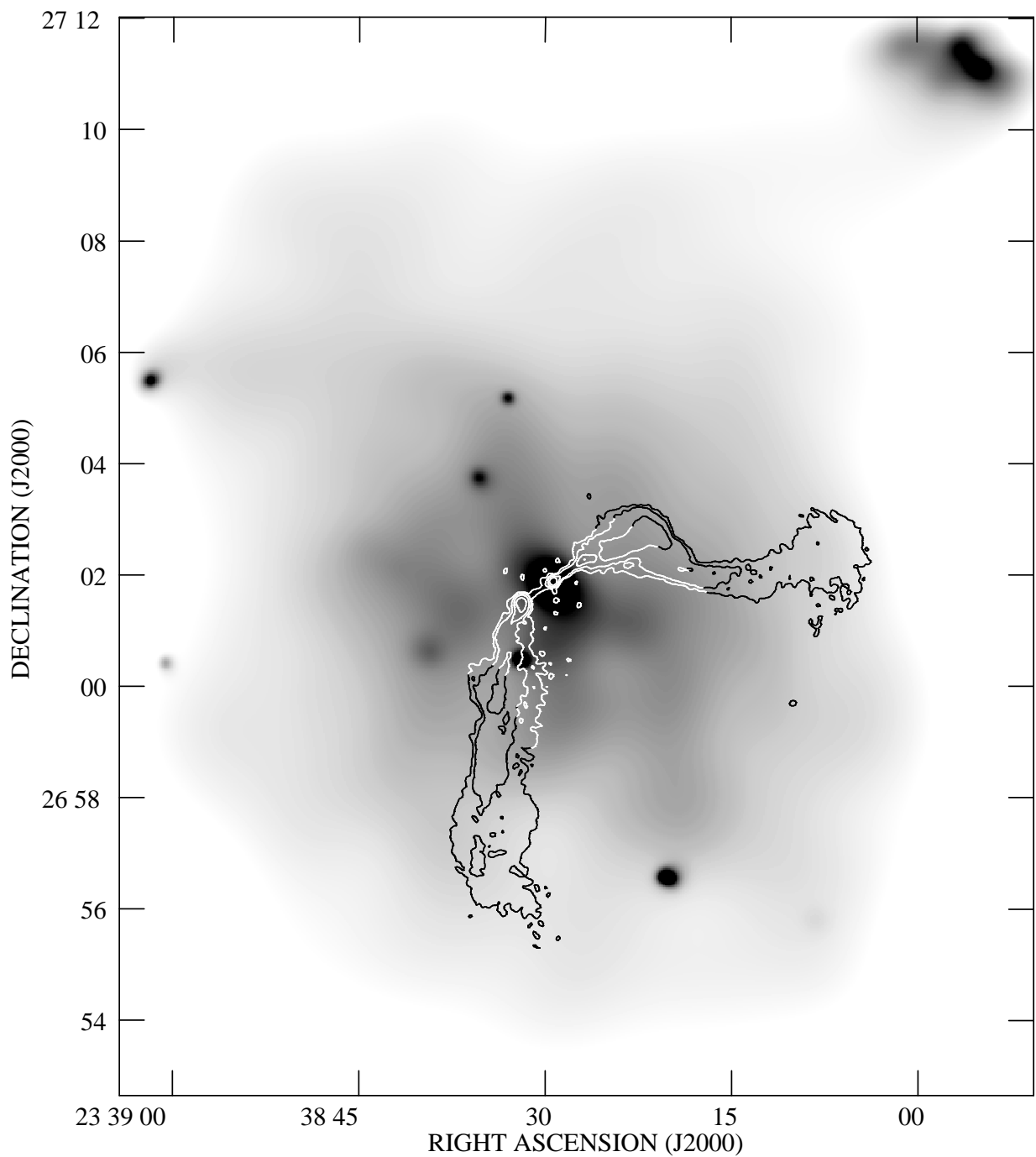}
\caption{Extended X-ray emission from Abell 2634. The greyscale shows
  the 0.5--5.0 keV data from the adaptively smoothed, combined field
  of view of all three {\it XMM} cameras after interpolation over chip
  gaps. The object to the NE is the background cluster CL-37.
  Superposed are contours of the 5.4-arcsec resolution 1.4-GHz VLA map
  at $1 \times (1, 4, 16\dots)$ mJy beam$^{-1}$. To avoid giving
  excessive weight to features at the edge of the field, this image is
  not vignetting-corrected.}
\label{xmm-image}
\end{figure*}

In the {\it Chandra} image there is a clear detection of an X-ray core
and emission associated with the jet, as well as small-scale extended
emission (Fig.\ \ref{chandra-image}). On larger scales, thermal
emission from the host cluster is clearly visible in the {\it Chandra}
data and there are detections of several of the individual galaxies of
the cluster. No significant excess X-ray emission is seen around the
points where the narrow inner jets enter the radio plumes. The {\it
XMM} observations (Fig.\ \ref{xmm-image}) detect the X-ray core and
small-scale thermal emission and the thermal emission from the
cluster, as well as some of the cluster galaxies and the background cluster
known as CL-37 (Scodeggio \etal\ 1995). In the following sections we
discuss each component in turn.

\begin{figure*}
\epsfxsize 8.5cm
\epsfbox{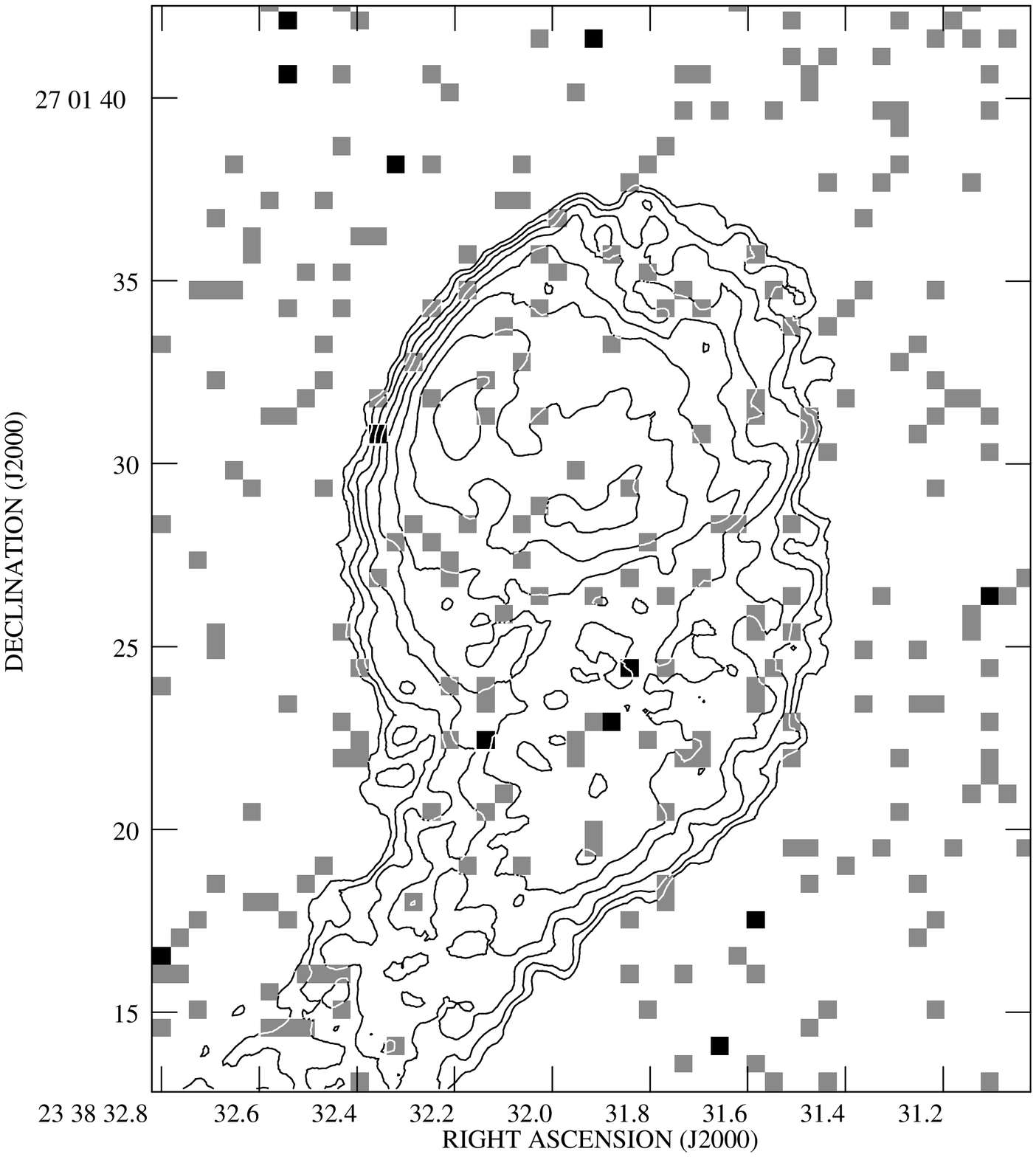}
\epsfxsize 8.5cm
\epsfbox{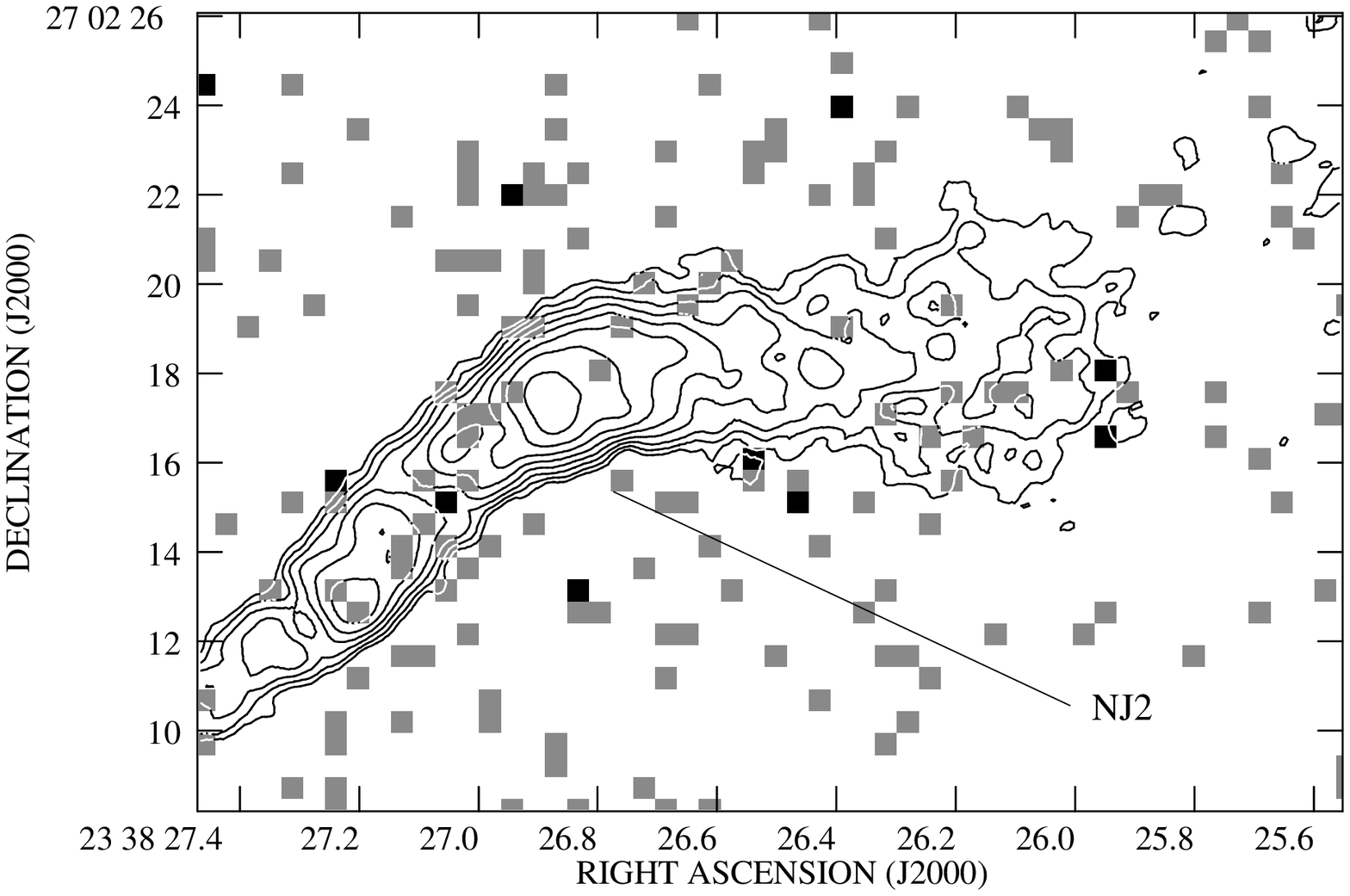}
\caption{X-rays near the plume bases of 3C\,465. Left: the south plume
base. Right: the north plume base. Pixels are 0.492 arcsec on a side: black
is 2 counts pixel$^{-1}$. Superposed are contours from a $0.72 \times
0.62$ arcsec resolution map at $0.1 \times (1, \sqrt{2}, 2, 2\sqrt{2}
\dots)$ mJy beam$^{-1}$. No X-ray emission is detected from any of the
bright features at the plume bases. The brightest knot, NJ2 in the
notation of Hardcastle \& Sakelliou (2004), is labelled.}
\label{pbase}
\end{figure*}

\subsection{The core}
\label{core}

There is small-scale thermal emission associated with the inner parts
of NGC 7720 (Section \ref{small-sc}), and so measurements of the
nuclear X-ray emission need to be carried out with care. We initially
used a small region (a circle of radius 1.5 arcsec centred on the peak
of the X-ray emission) to extract a spectrum from the {\it Chandra}
dataset. Background was taken from a concentric annulus between 1.5
and 2.5 arcsec. This spectrum was very poorly fitted ($\chi^2 = 159$ for
28 degrees of freedom) with a single power law model with Galactic
absorption, and the strong residuals below 1 keV suggested that there
was some contribution from thermal emission even in this small region.
A model consisting of a thermal ({\sc mekal}) component with 0.5 solar
abundance and an unabsorbed power law gave a much better fit ($\chi^2
= 48$ for 26 degrees of freedom), but required an unusually flat
photon index ($1.0 \pm 0.2$) for the power-law component. When the
power law was allowed to have some intrinsic absorption, a better fit
was obtained ($\chi^2 = 37.1$ for 25 d.o.f.): in this case the
temperature of the thermal component was $0.66 \pm 0.03$ keV, the
power-law photon index was $2.0_{-0.4}^{+0.5}$, and the excess
absorbing column was $5_{-2}^{+3} \times 10^{21}$ cm$^{-2}$. No
improvement in the fit was obtained by including further components,
such as a second unabsorbed power law. Imaging the core in hard and
soft bands (above and below 1 keV) suggests that it is indeed much
more point-like in the hard band than the soft band, supporting the
model implied by this fit. The power-law component contributes 260
counts in this model, or about 1/3 of the total in the extraction
region, and its unabsorbed 1-keV flux density is $14_{-7}^{+18}$ nJy,
taking into account a correction for the small aperture and local
background subtraction estimated from PSF simulations using the {\sc
chart} and {\sc marx} tools. Errors quoted here are $1\sigma$ for two
interesting parameters to take account of the strong relationship
between absorbing column and normalizing flux.

To compare the {\it Chandra} and {\it XMM} results, we extracted a
spectrum for both using a circle of radius 30 arcsec, taking
background from an annulus between 30 and 40 arcsec. (This region 
includes emission from the companion galaxy, NGC 7720A, which is 14
arcsec to the north, but as
discussed in Section \ref{compgs} this is a weak X-ray source and
does not contribute significantly to the fits.) The results are
similar to those found above: the best joint fit to the
thermal/absorbed power-law model, with all parameters the same for
both {\it XMM} and {\it Chandra} datasets, has $\chi^2 =134$ for 122
d.o.f. and gives a somewhat higher $kT = 0.86 \pm 0.02$ keV, an
intrinsic absorbing column for the power-law component
$3.7_{-1.3}^{+2.2} \times 10^{21}$ cm$^{-2}$, power-law photon index
$2.45_{-0.23}^{+0.27}$, and unabsorbed 1-keV power-law flux density of
$45_{-13}^{+20}$ nJy. The improvement in the fit obtained by allowing
the parameters of the power law component in the {\it Chandra} and
{\it XMM} datasets to vary independently is marginal ($\chi^2 = 126$
for 119 d.o.f.) and the best-fitting parameters are similar. A more
significant improvement ($\chi^2 = 120$ for 120 d.o.f.) is obtained by
fitting two {\sc mekal} models with different temperatures together
with the absorbed power law, suggesting that there may be temperature
structure in the extraction region: in this model the lower
temperature is $0.71 \pm 0.05$ keV, and so is consistent with the
temperature measured in the inner 1.5 arcsec, while the higher
temperature is $1.18_{-0.08}^{+0.13}$ keV. We discuss temperature
structure in more detail in Section \ref{small-sc}. The normalizations
for the power-law component in both these fits agree with that
determined from the small-aperture {\it Chandra} fit described above
within the large fitting uncertainties.

\subsection{The jet}

The {\it Chandra} detection of X-ray emission from the inner part of
the N jet of 3C\,465 is the first X-ray detection of a WAT jet. X-ray
emission appears to come predominantly from the two brightest knots of
the radio jet, denoted NJ1 by Hardcastle \& Sakelliou (2004) (Fig.
\ref{chandra-image}). It is hard to assign a statistical significance
to the detection of the inner knot (labelled knot 1 in Fig.\
\ref{chandra-image}), because of the difficulty in measuring a good
background so close to the bright, asymmetrical central X-ray emission
from the core and central thermal component: our best estimate, based
on local matched background circles, is that it contains $14 \pm 7$
counts. Knot 2, which corresponds to the brightest discrete feature of
the radio jet, contains $30 \pm 6$ counts, and there are a few more
counts plausibly associated with the jet in the bright jet region
beyond it. No X-ray emission from the jet is detected beyond 7.5
arcsec (4.4 kpc). Knot 2 is resolved transversely in the radio, with a
deconvolved Gaussian FWHM of 0.3 arcsec; this suggests that it should
be essentially unresolved to {\it Chandra}, and the observed X-ray
emission is consistent with that suggestion. There are too few counts
in the region of the jet from knot 2 onwards to extract a spectrum and
carry out detailed model fitting, but the two-bin spectrum that can be
extracted from the region marked on the figure suggests a steep X-ray
power-law photon index ($\sim 2.4$). Assuming this photon index and
Galactic absorption, the X-ray flux density from this region of the
jet is $0.9 \pm 0.14$ nJy. The corresponding radio flux density is 6.6
mJy at 8.4 GHz, so that the two-point radio-X-ray spectral index is
0.92. The unabsorbed jet flux in the {\it ROSAT} band is $10^{-14}$
ergs cm$^{-2}$ s$^{-1}$, corresponding to a luminosity of $2 \times
10^{40}$ ergs s$^{-1}$ at the distance of 3C\,465 (assuming no
beaming). The inner region of the jet is clearly different from the
outer region (beyond 7.5 arcsec): if the straight region of the outer
jet were similar in X-ray properties to the inner parts, we would
expect to have detected it at the level of $\sim 150$ {\it Chandra}
counts, whereas fewer than 30 ($3\sigma$) are detected. However, the
outer jet shows no clear knots of radio emission until the base of the
plume (see below).

\subsection{`Hotspots' and plume bases}

There is no significant detection of X-ray emission from the
radio-bright bases of the plumes in 3C\,465 (Fig. \ref{pbase}). The
strongest limit on the radio/X-ray ratio from these regions comes from
the bright southern plume base. In a circular region around the
brightest radio emission from this region we can set an upper limit of
25 {\it Chandra} counts, corresponding to $<0.6$ nJy assuming a
power-law model with a photon index of $\Gamma = 2.0$. The radio flux
density in the same region is 143 mJy, so that the two-point
radio-X-ray spectral index is $\alpha_{\rm RX} >1.12$. This is steeper
than is observed for any detected FRI jet base; if the plume base had
had a more typical value for FRI jets of $\alpha_{\rm RX} \approx
0.92$ (similar to the knots in the inner jet, see above), it would
have been detected at the level of $\sim 800$ counts. As pointed out
by Hardcastle \& Sakelliou (2004), 3C\,465 does not show any obvious
hotspot-like jet termination features, but the best candidate feature,
the knot at the end of the northern radio jet (NJ2 in their notation: labelled
in Fig.\ \ref{pbase}) is also
not significantly detected in the X-ray.

\subsection{Small-scale thermal emission}
\label{small-sc}

\begin{figure}
\epsfxsize \linewidth
\epsfbox{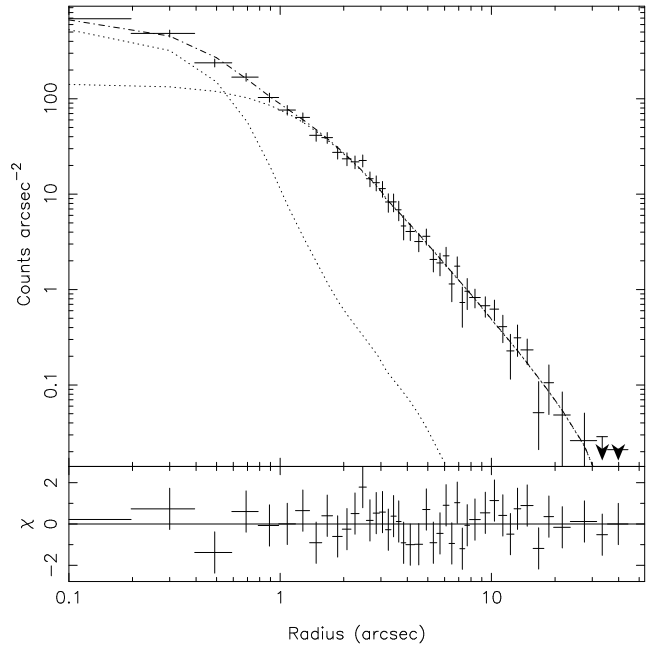}
\caption{Radial profile of the inner 35 arcsec of {\it Chandra} data
  in the energy range 0.5--5 keV. The dotted lines show the two
  best-fitting model components described in the text (point source
  and $\beta$ model) and the dashed line shows their sum. The
  residuals are displayed below the plot.}
\label{proplot}
\end{figure}

To characterize the small-scale thermal emission implied by the images
and the spectral fits of Section \ref{core}, we carried out radial
profile fitting to the inner 35 arcsec of the {\it Chandra} dataset,
taking background from the region between 35 and 45 arcsec and
excluding a pie slice at the position angle of the jet and the
emission from the nearby companion galaxy (see Section \ref{compgs}).
We use the PSF parametrization of Worrall \etal\ (2001).
The radial profile is best fitted with a combination of a small
$\beta$ model and a point-like component (Fig.\ \ref{proplot}). The
fit is good ($\chi^2 = 21.9$ for 40 d.o.f.) and the best-fitting model
parameters are $\beta = 0.62 \pm 0.03$, $\theta_c = 1.1 \pm 0.2$
arcsec (errors are $1\sigma$ for 2 interesting parameters). In this
model the point-like component contributes $400_{-30}^{+23}$ counts in
the 0.5--5 keV energy range: this compares well with the number
estimated from spectral fitting in Section \ref{core}. These
observations confirm the suggestion based on {\it ROSAT} data
(Sakelliou \& Merrifield 1999) that there is a component of extended
X-ray emission associated with NGC 7720.

We used spectral fitting in concentric annuli in the inner 30-arcsec
region to investigate whether there was evidence for temperature
structure in this small-scale thermal component. Identical background
regions, between 30 and 40 arcsec, were taken for each fit. The
results are shown in Table \ref{tgrad}. Either simply fitting spectra
to each annulus, or using the {\sc xspec} deprojection model, we find
evidence that there is a temperature gradient in the small-scale
thermal material in the centre of A2634.

Using the temperatures measured above in combination with the radial
profile, we estimate the central density of the small-scale thermal
material in 3C\,465 to be $(4.6 \pm 0.6) \times 10^5$ m$^{-3}$, while the
central pressure is $(1.1 \pm 0.2) \times 10^{-10}$ Pa. The central
cooling time of this material is short ($\sim 10^7$ years).
These conditions are very similar to those found in the central
components of more normal, non-WAT FRI radio galaxies hosted by groups
(e.g. Hardcastle \etal\ 2002, 2005).

\begin{table}
\caption{Temperatures in the central region of 3C\,465 measured with
  {\it Chandra}}
\label{tgrad}
\begin{tabular}{lrrrr}
\hline
Annulus&\multicolumn{2}{c}{Independent
  fits}&\multicolumn{2}{c}{Deprojected fits}\\
(arcsec)&Temperature&$\chi^2/n$&\multicolumn{2}{c}{Temperature (keV)}\\
&(keV)&(fixed $Z$)&(fixed $Z$)&(free $Z$)\\
\hline
0--1.5&$0.66_{-0.03}^{+0.03}$&37/25&$0.62_{-0.05}^{+0.04}$&$0.60_{-0.04}^{+0.04}$\\
1.5--3&$1.03_{-0.04}^{+0.04}$&27/16&$0.98_{-0.03}^{+0.03}$&$1.01_{-0.04}^{+0.04}$\\
3--7.5&$1.42_{-0.06}^{+0.12}$&32/17&$1.39_{-0.07}^{+0.06}$&$1.64_{-0.11}^{+0.10}$\\
7.5--30&$1.8_{-0.3}^{+0.8}$&9/13&$1.9_{-0.4}^{+1.0}$&$2.5_{-0.7}^{+1.2}$\\
\hline
\end{tabular}
\begin{minipage}{\linewidth}
The fit to the central bin includes an obscured power-law component as
discussed in Section \ref{core}. Abundance is fixed to 0.5 solar for
the temperatures listed in the `fitted' and `deprojected' columns. The
total $\chi^2/n$ values for the deprojected models were 109/71 (fixed
abundance) and 86/70 (free abundance). The best-fitting abundance for
the free-abundance deprojection model was $1.5_{-0.2}^{+0.4}$.
\end{minipage}
\end{table}

\subsection{The cluster galaxies}
\label{compgs}

\begin{figure}
\epsfxsize 8.8cm
\epsfbox{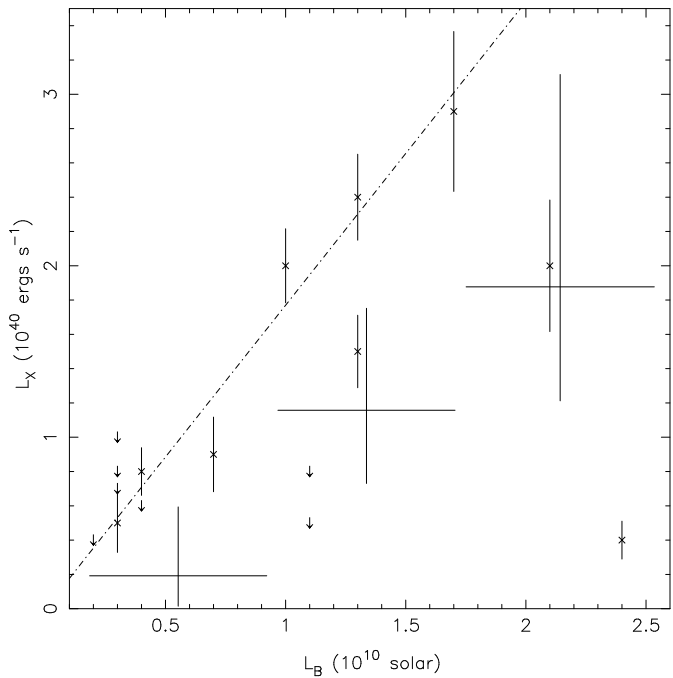}
\caption{X-ray luminosity versus blue optical luminosity for the
  early-type galaxies in the {\it Chandra} field of view. Data points
  are from Table \ref{galaxies}: the large crosses show the bins of
  Sakelliou \& Merrifield (1998) after correction to the cosmology we
  use in this paper, and the dotted line shows the boundary between
  binary- and thermal-dominated X-ray emission adopted by them (in the
  absence of a more up-to-date study of the relationship between X-ray
  and optical properties of elliptical galaxies using {\it Chandra}).}
\label{lxlb}
\end{figure}

Eleven galaxies from the A2634 cluster were detected in the {\it Chandra}
field of view; this is almost all of the bright cluster galaxies that
are suitably positioned for {\it Chandra} imaging. Of these, the
five brightest in X-rays and lying in the {\it XMM} field of view
were also detected with {\it XMM}. The detected sources are listed in
Table \ref{galaxies}, along with some less bright elliptical galaxies
that in general are not detected. We measured the {\it Chandra} counts
for each source using circular extraction regions and concentric or
adjacent background annuli, and, where spectral fitting was possible,
fitted each {\it Chandra} spectrum ({\it XMM} spectra gave little
additional information) with a combination of an unabsorbed, hard
power-law model with fixed $\Gamma = 1.0$ and a {\sc mekal} model with
fixed 0.5 solar abundance (the power-law component being intended to
represent the contribution from X-ray binaries). Where galaxies were
not detected, we set upper limits based on the Chandra background.
Results are tabulated in Table \ref{galaxies}. Where spectral fits
were possible, most galaxies were fitted with a thermal model with a
temperature between 0.5 and 0.8 keV, in some cases together with a
less luminous power-law component. Cluster galaxies therefore do
contain some hot gas, though the luminosity even of the brightest
system is about an order of magnitude less than that of the extended
X-ray emission associated with NGC 7720 itself. The typical galaxy
X-ray luminosities -- calculated from the fitted model where possible,
and using a thermal model with $kT = 0.5$ keV for the weaker sources
or upper limits -- are comparable to the luminosities estimated for
the class of optically brighter galaxies by Sakelliou \& Merrifield
(1998), after correction for the different cosmology used in that
paper; this suggests that the analysis carried out in that paper,
where individual galaxies were not significantly detected, was broadly
correct, although the presence of some emission best modelled as
thermal suggests that the galaxy atmospheres have not been fully
stripped. When plotted on the $L_X/L_B$ relation (Fig.\ \ref{lxlb}),
some of the early-type galaxies we detect lie above the relation found
by Sakelliou \& Merrifield (their fig. 6), though the errors on their
relation are large. However, it should be noted that several of the
galaxies we detect are unusual in the sense of being possibly
interacting (e.g. P257 and 259) and were excluded from the analysis of
Sakelliou \& Merrfield.

\begin{table*}
\caption{Galaxies detected in the {\it Chandra} observation and upper
  limits on non-detections.}
\label{galaxies}
\begin{tabular}{lllllrrlrrr}
\hline
Galaxy ID&Type&{\it XMM}&RA&Dec&Ext'n&{\it Chandra}&Spectral fit&$\chi^2/n$&Luminosity&$L_B$\\
&&det'n?&h m s&$^\circ$ $'$ $''$&radius&counts&&&($\times
10^{40}$&($\times 10^{10}$\\
&&&&&(arcsec)&&&& ergs s$^{-1}$)&$L_{\sun}$)\\
\hline
P202&E&U&23 38 29.6&27 02 05&3&$31 \pm 7$&--&&0.5\\
P246&S0&N&23 38 52.5&26 56 22&15&$<21$*&&&$<0.7$&0.3\\
P257&E&Y&23 38 29.2& 26 58 44&9&$134 \pm 14$&{\sc mekal} + pow, $kT =
0.8 \pm 0.1$&1.1/3&2.4&1.3\\
P256&E&N&23 38 34.4& 26 58 46&9&$69 \pm 12$&{\sc mekal}, $kT = 0.8_{-0.15}^{+0.75}$&1.4/1&0.8&0.4\\
P259&E&Y&23 38 26.8& 26 59 06&8&$78 \pm 11$&{\sc mekal} + pow, $kT = 0.5
\pm 0.2$&0.5/1&1.5&1.3\\
P263&E&Y&23 38 38.8&27 00 41&9&$139 \pm 15$&{\sc mekal}, $kT = 0.61 \pm 0.05$&4.0/4&2.0&1.0\\
P271&S0&N&23 38 36.3& 27 01 48&9&$62 \pm 15$&--&&0.9&0.7\\
P275&E&N&23 38 33.3&27 02 05&5&$29 \pm 8$&--&&0.4&2.4\\
P294&S0&N&23 37 53.7&27 05 18.4&20&$<30$*&&&$<1.0$&0.3\\
P295&S0&N&23 38 27.1&27 05 25&10&$32 \pm 11$&--&&0.5&0.3\\
P302&S&Y&23 39 11.7& 27 06 55&19&$84 \pm 12$*&pow&2.8/3&2.7\\
P308&E&N&23 38 22.8&27 09 29&10&$<33$&&&$<0.5$&1.1\\
P316&S0&N&23 38 46.3&27 10 20&15&$<20$*&&&$<0.6$&0.4\\
P320&S0&N&23 39 07.0&27 12 21&20&$<24$*&&&$<0.8$&0.3\\
P321&S0&N&23 38 41.7&27 12 54&10&$<13$*&&&$<0.4$&0.2\\
P322&S0&N&23 38 50.1&27 12 53&20&$<26$*&&&$<0.8$&1.1\\
P324&S0&Y&23 38 43.7& 27 12 57&22&$87 \pm 14$*&{\sc mekal}, $kT = 0.32_{-0.04}^{+0.1}$&2.8/2&2.9&1.7\\
P331&E/S0&--&23 38 50.5& 27 16 06&23&$73 \pm 14$*&{\sc mekal}, $kT = 0.6\pm0.1$&0.3/1&2.0&2.1\\
\hline
\end{tabular}
\begin{minipage}{18cm}
The galaxy IDs refer to the numbers allocated by Pinkney \etal\
(1993): P202 is the companion galaxy, also known as NGC 7720A. The
{\it XMM} detection column indicates whether a galaxy was detected by
{\it XMM} (Y), undetected although in field of view (N) or
undetectable due to proximity to another source (U) or through being
out of the field of view (--). Extraction regions used reflect the
detected X-ray emission, the need to avoid including emission from
other sources, and the size of the PSF at the off-axis distance of the
source: the standard extraction region of 9 arcsec for the on-axis
sources corresponds to 5 kpc at the distance of the cluster. {\it
  Chandra} counts quoted are in the 0.5--5 keV energy range. Counts
marked with an asterisk are from one of the front-illuminated chips,
the others are from ACIS-S3. X-ray luminosities are in the {\it
  Einstein} range (0.2--3.5 keV) and are calculated from the absorbed
flux assuming a luminosity distance of $3.96 \times 10^{26}$ cm. Where
no spectral fit was possible, a {\sc mekal} model with $kT = 0.5$ keV
was used to estimate luminosity. Optical luminosities are taken from
the values used by Sakelliou \& Merrifield (1998), updated to the
cosmology of the present paper.
\end{minipage}
\end{table*}

\subsection{The cluster}
\label{cluster}

\subsubsection{Cluster structure}
\label{wake}

\begin{figure*}
\epsfxsize 8.8cm
\epsfbox{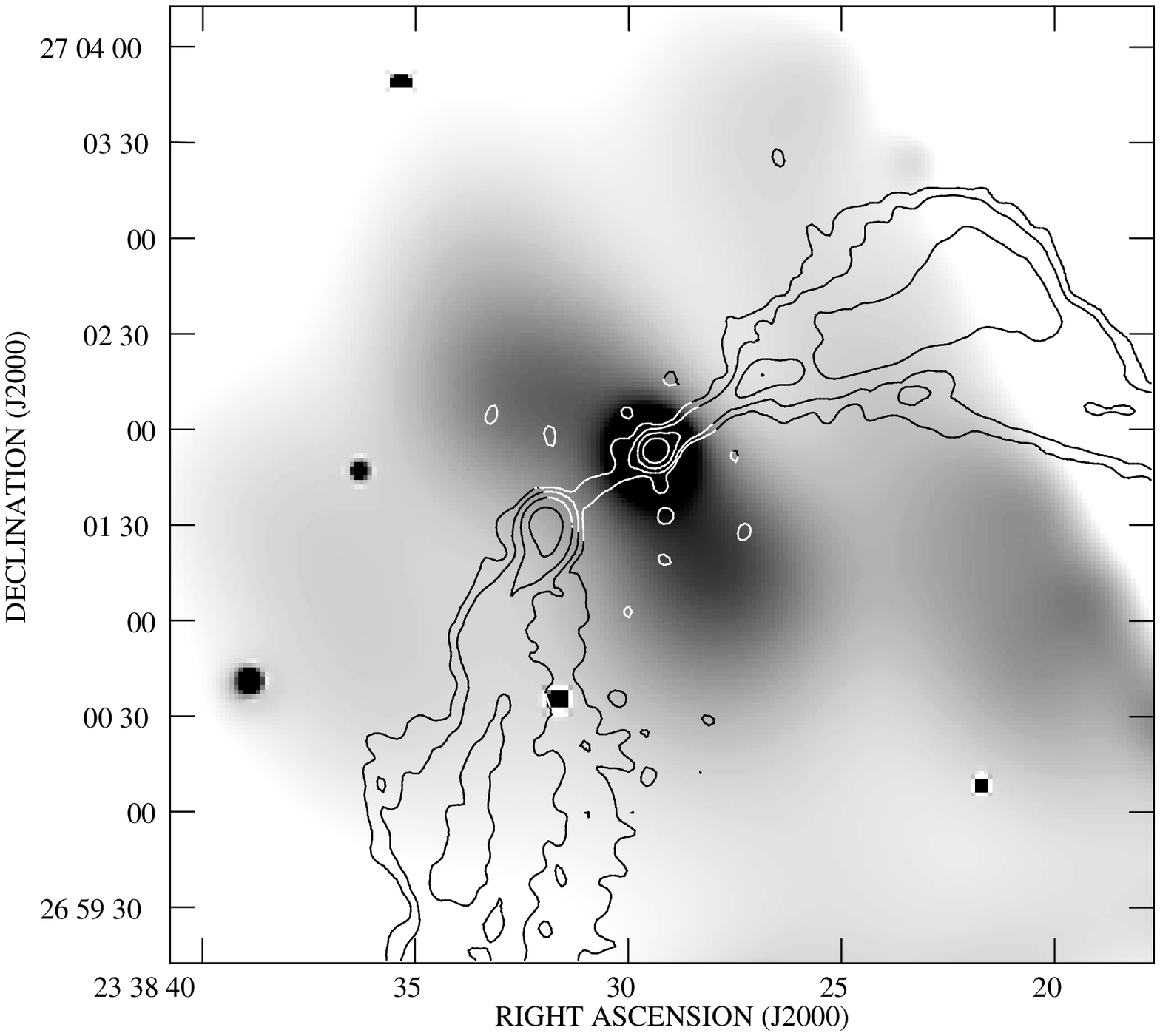}
\epsfxsize 8.8cm
\epsfbox{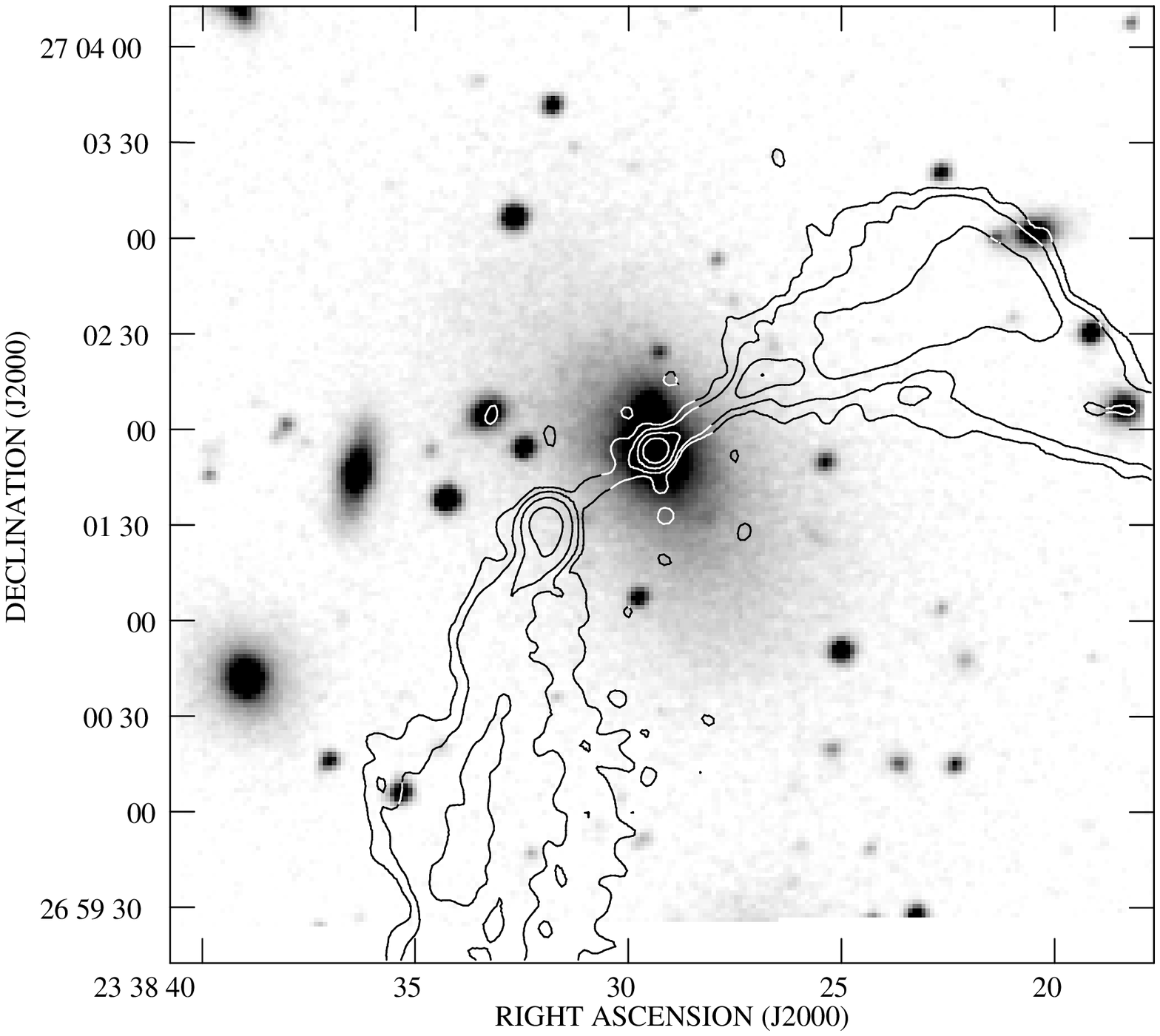}
\caption{The inner parts of the 3C\,465 plumes superposed on (left)
  the adaptively smoothed image from the ACIS-S3 chip of {\it Chandra}
  and (right) the Digital Sky Survey 2 red image of the host galaxy and
  its environment. Radio map and contours as in Fig.\ \ref{xmm-image}.
  The galaxies to the E, with corresponding X-ray sources, are P271
  and P263 in the notation of Table \ref{galaxies}.}
\label{chadapt}
\end{figure*}

Examination of the image shown in Fig.\ \ref{xmm-image} shows that
there is considerable structure in the large-scale cluster emission,
as inferred previously from the {\it ROSAT} images. We have verified,
by masking out the {\it XMM} and {\it Chandra}-detected point sources
using the {\it dmfilth} tool in {\sc ciao},
that little of the structure seen in Fig.\ \ref{xmm-image} can be
attributed to combinations of weak point sources. Three features of
the image are of particular interest:

\begin{enumerate}
\item The elongation of the central X-ray emission on arcmin scales in
  a direction perpendicular to that of the jet axis, but similar to
  that of the isophotes of the host galaxy (features that were
  originally pointed out by E84). The elongation can be
  seen particularly clearly in the {\it Chandra} data, as illustrated
  in Fig.\ \ref{chadapt}.
\item The larger-scale elongation of the X-ray emission in the same
  direction, particularly to the SW. This is the extension seen in
  the images of Schindler \& Prieto (1997) and Sakelliou \& Merrifield (1999).
\item The apparent decrements in X-ray surface brightness at the
  positions of the bases of the plumes. Measurements of count density
  using regions defined from the adaptively smoothed maps shows that
  these deficits are significant at the $\sim 2$--$3\sigma$ level. (It
  should be noted, however, that there are apparent deficits at
  similar significance to the NE and SW, which do not correspond to
  the positions of any radio feature.)
\end{enumerate}

\begin{figure}
\epsfxsize 8.8cm
\epsfbox{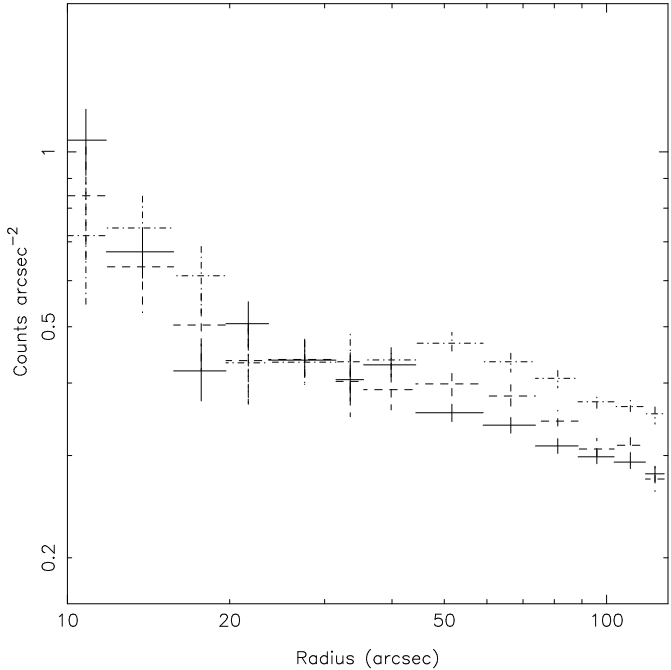}
\caption{Radial profile of the {\it Chandra} data in the 0.5--5 keV
  energy range. Dot-dashed lines show the surface brightness profile
  in the quadrant centred on position angle 205$^\circ$ (N through E),
  or `behind' the galaxy: dashed lines show it in the quadrant centred
  on PA=50$^\circ$, and solid lines show the profile in the remaining
  180$^\circ$. No background subtraction has been carried out. A
  significant difference is apparent from a radius of about 40 arcsec
  outwards.}
\label{angprofile}
\end{figure}

It is worth commenting briefly on the elongation of the X-ray emission.
Fig.\ \ref{chadapt} strongly suggests that the extension is brighter
in the SW direction, or `behind' the galaxy if we were to adopt a
model in which the plume bending is related to the motion of the
galaxy (see Section \ref{bending} for discussion of this point). To
confirm this, we carried out radial profiling in angular ranges guided
by the structure seen in Fig. 6. The results are shown in Fig.\
\ref{angprofile}; a clear excess in both directions can be seen, but
the extension in the SW direction is systematically brighter. This
agrees with the results that E84 found on larger scales. They
commented that the optical galaxy shows the same asymmetry, a result
we confirm by identical radial profiling of the DSS-2 image. In a
model in which the host galaxy is moving through the ICM, it is
tempting to model the excess extension in the SW direction as a `wake'
due to ram-pressure stripping of the small-scale thermal component
and/or Bondi-Hoyle accretion behind the galaxy. At least one other WAT
exhibits X-ray evidence for a wake (Sakelliou \etal\ 1996, 2005).
While we do not rule out the possibility of a weak wake in 3C\,465,
the wake model leaves two questions unanswered: 1) how can
ram-pressure and/or accretion explain the asymmetry in the {\it stars}
of the host galaxy? and 2) how does the small-scale wake relate to the
elongation on much larger scales, which extends both to the NE and the
SW of the present position of the host galaxy?

\subsubsection{Global properties}

\begin{figure}
\epsfxsize \linewidth
\epsfbox{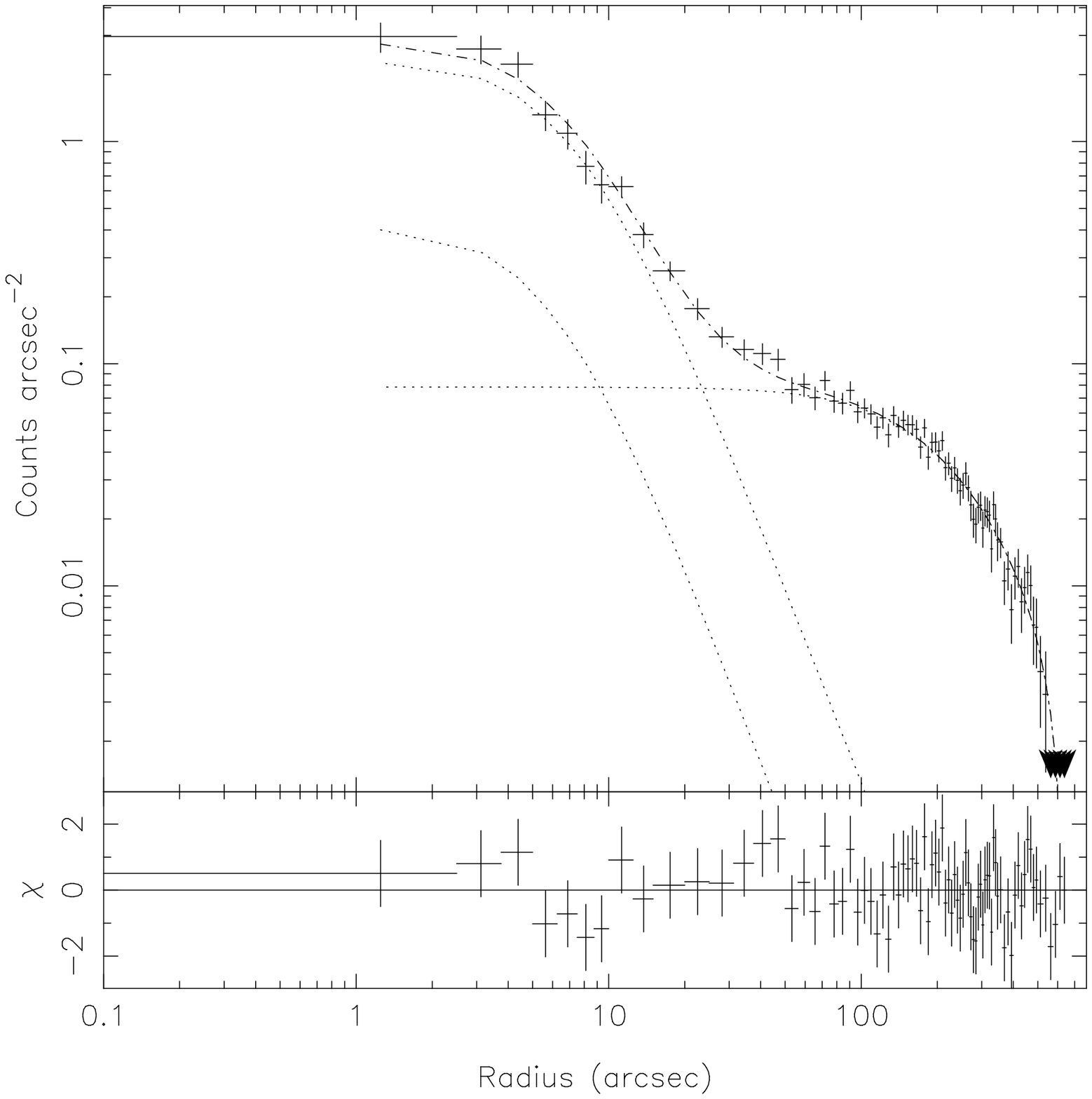}
\caption{Radial profile of the {\it XMM} pn data in the energy range
  0.3--7 keV. The dotted lines show the three best-fitting model
  components described in the text (point source and two $\beta$
  models) and the dashed line shows their sum. The residuals are
  displayed below the plot.}
\label{pnrprof}
\end{figure}

To characterise the spatial and spectral structure of the extended
emission we used the same techniques as we applied to the small-scale
structure: as Schindler \& Prieto (1997) argued, the departures from
radial symmetry are predominantly on smaller scales and in any event
do not dominate in terms of counts. We therefore extracted radial
profiles for each camera in annuli centred on the X-ray core and
extending to a radius of 625 arcsec, carrying out background
subtraction in the way described by Croston \etal\ (2003). (Unlike our
spectral extractions, the radial profile extraction {\it does} use
double subtraction to take account of any differences between the
cosmic X-ray background in our observation and in the background
files: this is possible in this case because the radial profile
fitting code we use can take into account the contribution of the
fitted model to the background region.) The model
fitted consisted of a point component, a $\beta$ model whose
structural parameters were fixed to the best-fitting parameters for
the small-scale emission determined with {\it Chandra} (Section
\ref{small-sc}) and a $\beta$ model whose core radius and $\beta$
value were allowed to vary. The normalization of each component was
allowed to vary. The best-fitting large-scale $\beta$ model to the
profiles from all three cameras had $\beta = 0.41 \pm 0.06$, $\theta_c
= 180 \pm 40$ arcsec (errors are $1\sigma$ for two interesting
parameters) with $\chi^2 = 201$ for 237 degrees of freedom. The fit to
the pn data is shown in Fig.\ \ref{pnrprof}; fits to the MOS profiles
are similar. The fits agree with the {\it Chandra} fits in suggesting
that the small-scale extended component dominates over the
point-source component, though the uncertainties on the relative
normalizations of the two components are large: the total counts found
by the fits in the two small-scale components are consistent with what
would be expected given the count rate obtained from the {\it Chandra}
data.

The spectral properties of various regions of the cluster were
estimated using the {\sc xspec} deprojection model in combination with
{\it mekal} thermal models and a power-law contribution in the central
region. To make this model work effectively the regions used must
cover a large region of the cluster, and so we used annuli extending
out to 10 arcmin (as tabulated in Table \ref{xmm-annuli}) with
background taken from the background files and with the corrections
described in Section \ref{xmm-intro}. As with the {\it Chandra} fits
described above, we also tabulate the results of fitting independently
to the annuli used. The abundance was initially fixed at 0.5 solar: if
it was allowed to vary, the best-fitting joint value in the
deprojection model was $0.46 \pm 0.06$, with very similar temperatures
(not tabulated).

\begin{table}
\caption{Temperatures in A2634 measured with {\it XMM}}
\label{xmm-annuli}
\begin{tabular}{lrrr}
\hline
Annulus&\multicolumn{2}{c}{Independent
  fits}&\multicolumn{1}{c}{Deprojected fits}\\
(arcsec)&Temperature&$\chi^2/n$&Temperature\\
&(keV)&&(keV)\\
\hline
0--30&$0.82_{-0.04}^{+0.03}$&75/71&$0.91_{-0.04}^{+0.03}$\\
30--60&$4.3_{-0.4}^{+0.4}$&98/76&$2.9_{-0.6}^{+1.5}$\\
60--120&$4.8_{-0.3}^{+0.3}$&217/221&$4.3_{-0.8}^{+0.8}$\\
120--240&$5.1_{-0.3}^{+0.3}$&323/280&$5.2_{-1.5}^{+1.8}$\\
180--240&$5.0_{-0.3}^{+0.3}$&315/335&$6.5_{-1.0}^{+2.7}$\\
240--300&$4.8_{-0.3}^{+0.3}$&329/372&$5.6_{-1.4}^{+4.1}$\\
300--600&$4.6_{-0.1}^{+0.2}$&318/475&$4.5_{-0.1}^{+0.1}$\\
\hline
\end{tabular}
\begin{minipage}{\linewidth}
The fit to the central bin includes an obscured power-law component as
discussed in Section \ref{core}. Abundance is fixed to 0.5 solar (see
the text). The total $\chi^2/n$ value for the deprojected model was
1671/1847.
\end{minipage}
\end{table}

It will be seen that the deprojected and independent fits give similar
temperatures, around 4.5--5 keV, for the outer regions of the cluster.
These temperatures are in reasonable agrement, within the errors, with
the temperatures estimated from {\it ROSAT} PSPC data by Schindler \&
Prieto (1997), though somewhat higher than the cluster temperature of
$3.7 \pm 0.3$ keV derived from {\it ASCA} data by Fukazawa \etal\
(1998). It remains possible that the temperatures we measure are
contaminated by excess particle background, in spite of the steps we
have taken to subtract this off (Section \ref{xmm-intro}). However, we
see no sign of an increase in temperature at large radii, as we might
expect in such a scenario (and as we do in fact see without the
particle background corrections). Instead, the temperatures are
largely consistent with a constant value of $\sim 4.7$ keV beyond
about 1 arcmin from the core.

\begin{figure}
\epsfxsize \linewidth
\epsfbox{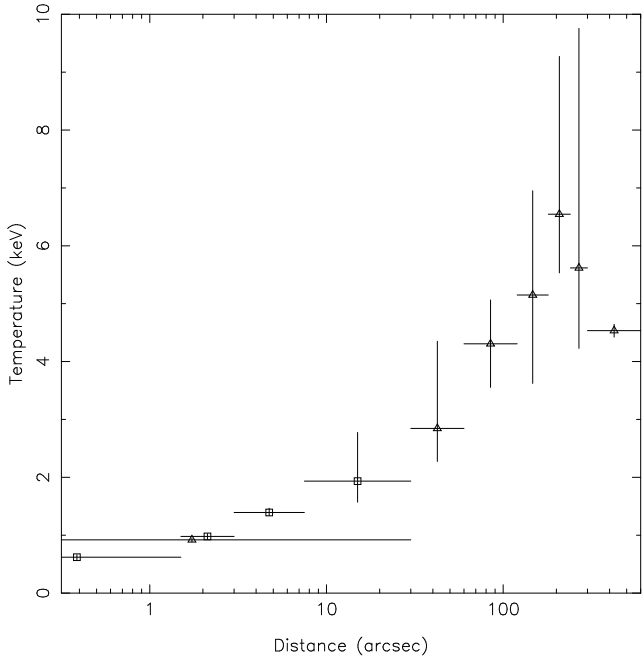}
\caption{The deprojected temperature of the thermal emission
  surrounding 3C\,465, based on {\it Chandra} and {\it XMM}
  deprojections (Tables \ref{tgrad} and \ref{xmm-annuli}). In the
  inner 30 arcsec data are plotted from both {\it XMM} (one large
  region) and {\it Chandra} (several small regions). Squares indicate
  {\it Chandra} data and triangles {\it XMM} data points.}
\label{tprofile}
\end{figure}

Combining the {\it XMM} deprojection with the {\it Chandra}
deprojection of Section \ref{small-sc}, we obtain the temperature
profile shown in Fig.\ \ref{tprofile}. Whereas the {\it XMM} data
alone might have suggested a significant temperature jump at around
the jet termination distance in 3C\,465 ($\sim 40$ arcsec) the two
datasets taken together strongly suggest a smooth temperature
gradient as a function of distance, probably levelling out at $kT \sim
4.7$ keV. 

\begin{figure}
\epsfxsize \linewidth
\epsfbox{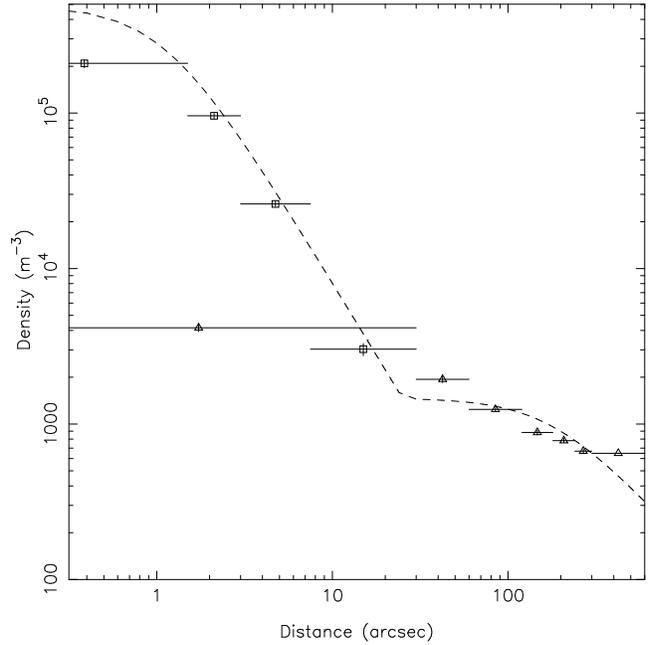}
\caption{The deprojected proton density of the thermal emission
  surrounding 3C\,465, based on {\it Chandra} and {\it XMM}
  deprojections (Tables \ref{tgrad} and \ref{xmm-annuli}) (solid
  bars). Overplotted are the densities estimated from radial profile
  fitting of $\beta$ models (dashed lines), assuming a transition
  between the two $\beta$ models at 20 arcsec from the core. Squares indicate
  {\it Chandra} data and triangles {\it XMM} data points.}
\label{dprofile}
\end{figure}

The deprojection allows us to estimate particle densities in the
deprojection annuli, which we can compare with those estimated from
the radial profile analysis using the method of Birkinshaw \& Worrall
(1993). The radial profile values are necessarily calculated assuming
a single conversion factor between count density and volume-normalized
emission measure, and so cannot take account of temperature structure:
we carry out the conversion assuming instrumental responses
appropriate for temperatures of 4.7 keV for the outer region modelled
with {\it XMM} data and 1.0 keV for the {\it Chandra} data that
describe the inner region. On the other hand, the densities estimated
using the deprojection model are emission-weighted densities and can
only be at best representative of the density in the bin. The proton
densities estimated using the two models are plotted as a function of
radius in Fig. \ref{dprofile} (we assume for the purposes of the plot
that the transition between the two $\beta$ models happens when their
densities become equal at around 20 arcsec from the core). The plot
shows that the agreement between the densities estimated using the two
methods is reasonable, except perhaps in the outer bin, which would be
expected to be most affected by the assumption of the deprojection
algorithm that no cluster emission exists outside the deprojection
region. We can therefore have reasonable confidence in pressures
estimated using either method. The agreement between the two methods
also suggests that the physical conditions we estimate are reliable in
spite of the fact that we find lower $\beta$ values and core radii
than other workers.

\begin{figure}
\epsfxsize \linewidth
\epsfbox{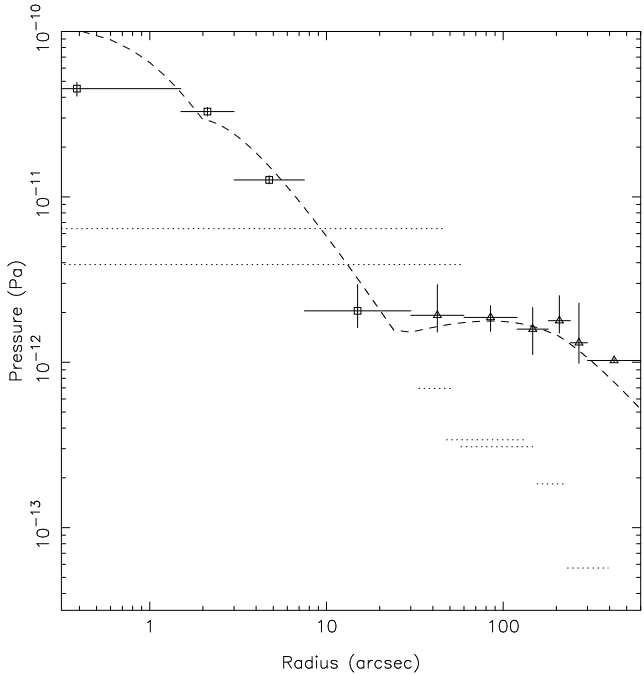}
\caption{The pressure of the thermal emission surrounding 3C\,465,
  based on deprojections as plotted in Figs \ref{tprofile} and
  \ref{dprofile} (solid bars). The dashed lines are the pressures
  estimated from the radial profile density determinations, converted
  to pressure by assuming a temperature profile as described in the
  text. Dotted lines show the synchrotron minimum pressures in various
  components of 3C\,465, with the inner and outer radii corresponding
  to the inner and outer positions of the relevant component. The two
  inner lines show the minimum pressure in the jet 1) assuming no
  beaming in the jet and a jet in the plane of the sky and 2) with a
  representative beaming model with $v/c = 0.5$, $\theta = 50^\circ$.
  All other calculations assume no beaming or projection. Squares
  indicate {\it Chandra} data and triangles {\it XMM} data points.}
\label{pprofile}
\end{figure}

Pressure as a function of radius, estimated from the deprojection and
from the radial profile fitting, is plotted in Fig.\ \ref{pprofile}.
To convert the density estimated from radial profile fitting to a
pressure we have assumed a temperature which is constant at 0.66 keV
within 2 arcsec and constant at 4.7 keV beyond 200 arcsec, and has a
log-linear profile between those two radii (i.e. $kT = a + b \log r$):
such a temperature profile is a reasonable description of the profile
plotted in Fig. \ref{tprofile}, though others could be devised. It can
be seen that the pressure estimates are again reasonably consistent,
as expected. Also plotted on Fig.\ \ref{pprofile} (dashed lines) are
the minimum pressures in various components of the radio source. These
are calculated from flux measurements from the 1.4 GHz and 8.4 GHz
maps on the assumption of a synchrotron spectrum with $\gamma_{\rm
min} = 100$, $\gamma_{\rm max} = 10^5$, a low-energy electron energy
index of 2 (corresponding to $\alpha = 0.5$) and a break in the electron
spectrum determined by the fluxes in the two maps. We assume no
relativistic protons, a filling factor of unity, and a cylindrical
geometry for all features. For the bright northern jet, we have
estimated the minimum pressure in two ways: firstly assuming no
projection or beaming (as was the case for all the other components)
and secondly assuming an effective beaming speed of $0.5c$ and an
angle to the line of sight $\theta=50^\circ$: this latter is a
representative beaming value that reproduces the jet sidedness ratio
measured by Hardcastle \& Sakelliou (2004). These pressures are
essentially true minimum pressures: only very implausible assumptions
for the electron energy spectra can reduce them significantly, and then only by
a factor $<2$. Fig.\ \ref{pprofile} shows
that the inner jet is the only component, on either of these two
models, that appears to have a higher minimum internal pressure than
the external thermal material over some of its length. All the other
regions of the source are underpressured with respect to the external
pressure by factors of between about 3 and 30, as is typically found
for the large-scale components of FRI sources (Hardcastle \& Worrall
2000 and references therein). This conclusion is different from that
of E84 partly because our estimated minimum pressures
are slightly lower, but mostly because our external pressures are
higher, as a result of a higher measured external temperature (their
value of 1.85 keV is lower than any modern estimate) and a
better-constrained model for the external density.

\subsection{CL-37}

Pinkney \etal\ (1993) and Scodeggio \etal\ (1995) both report the
existence of a cluster of galaxies 12 arcmin to the NW of NGC 7720,
with a redshift around 0.124. This cluster is detected in X-rays in
the MOS1 and MOS2 cameras of {\it XMM} (it lies on the very edge of
the pn field of view). Extracting a spectrum from a 75-arcsec radius
circle around the X-ray centroid, with background taken from an
adjacent larger circle, we find that the MOS1 and MOS2 data are well
fitted ($\chi^2$/d.o.f. = 84/88) by a {\sc mekal} model with $kT = 3.1
\pm 0.4$ keV and abundance $0.4_{-0.2}^{+0.3}$ solar. The unabsorbed
X-ray flux in this region is around $10^{-12}$ ergs cm$^{-2}$
s$^{-1}$, which is substantially lower than the value predicted by
Scodeggio \etal\ from the luminosity-velocity dispersion relationship
even after correcting to the cosmology used here, although the
extraction region is not large at the distance of CL-37. However, as
the measured temperature would correspond (e.g. Edge \& Stewart 1991)
to an expected velocity dispersion in the range 400--600 km s$^{-1}$,
it seems likely that the true velocity dispersion of the cluster is
less than the value of $924_{-265}^{+307}$ km s$^{-1}$ quoted by
Scodeggio \etal . Given the large uncertainties on the result of
Scodeggio \etal\ , their comment that the velocity structure of the
cluster is complex, and the large scatter in the $kT$-$\sigma$
relationship, the true velocity dispersion can be estimated to be
something closer to $\sigma \approx 600$ km s$^{-1}$.

\section{Discussion}

\subsection{X-ray emission from the jet and plumes}

It is conventional to believe (e.g. Leahy 1993) that the jets of WAT
sources are physically similar to those in classical double (FRII)
radio galaxies and quasars; evidence for this comes from their narrow
and sometimes one-sided appearance (as in 3C\,465), the polarization
structure they exhibit and their frequent termination in compact,
shock-like features similar to the hotspots in FRIIs (Hardcastle \&
Sakelliou 2004). X-ray emission from FRII-like (`type II') jets,
particularly in quasars and other sources where the jet is {\it a
  priori} thought to make a small angle to the line of sight, is
often attributed in the literature to inverse-Compton scattering of
the cosmic microwave background radiation by a jet with a high bulk
Lorentz factor ($\Gamma \sim 10$), based on the observations that a
one-zone synchrotron model cannot be fitted to the data and that the
inverse-Compton process without beaming requires a large departure
from the equipartition condition (e.g. Tavecchio \etal\ 2000).
However, the generally accepted picture of X-ray emission from
FRI-like (`type I') jets is that they are due to synchrotron emission
related to the strong deceleration of the jet flow expected on other
grounds in the inner few kpc of these systems (e.g. Hardcastle \etal\
2001, 2002); in these cases a one-zone synchrotron model is often a
good fit to the overall jet spectrum (although the details of jet
structure may in fact be complex on scales that cannot be resolved in
most sources: Hardcastle \etal\ 2003) while a beamed inverse-Compton
model requires implausible jet parameters.

The detection of 3C\,465's inner jet in X-rays, at a level similar to
what would be predicted using a simple synchrotron model based on
observations of type I jets, lends support to the idea that some type
II jets can be synchrotron sources. There is already evidence for this
from the comparatively strong detections of the pre-flaring jets,
thought to be type II, in some FRI sources, notably 3C\,66B and Cen A
(Hardcastle \etal\ 2001, 2003) as well as X-ray detections of
components of the jets of some FRII sources (e.g. 3C\,219, Comastri
\etal\ 2003; 3C\,452 and 3C\,321, Hardcastle \etal\ 2004; 3C\,403,
Kraft \etal\ 2005) for which it is highly implausible that a beamed
inverse-Compton model can be an explanation. 3C\,465 is particularly
interesting in this context for three reasons: (1) it is generally accepted
that WAT jets are type II, (2) the emission from the jet is not
confused by inverse-Compton emission from any surrounding lobes, and
(3) there is no confusion between jet knots and hotspots, which can
also be synchrotron X-ray sources. 3C\,465's detection thus gives us a
reason to reiterate a point already made by Hardcastle \etal\ (2004)
and Kraft \etal\ (2005): there are clearly X-ray-emitting type II jets
for which the beamed inverse-Compton model is {\it not} the
explanation. Since there is no compelling reason to expect a one-zone
synchrotron model to be a good fit in an FRII jet in which the loss
spatial scale may be two to three orders of magnitude smaller than the
region of the jet being considered, observers should be wary of ruling
out the possibility that synchrotron emission is responsible for the
X-rays from a given type II jet.

On the other hand, the non-detection of the bases of the plumes of
3C\,465 in X-rays, at a level sufficient to show that their radio to
X-ray spectral index is significantly greater than that seen in all
FRI jets, strongly suggests that these regions are {\it not} analogous
to the bases of type I jets. This is as expected in the picture of
these objects proposed by Hardcastle \etal\ (1998) and Hardcastle \&
Sakelliou (2004), in which the jet decelerates rapidly at a single
point close to the base of the plume, rather than there being uniform
distributed deceleration over a finite-sized region of the base of the
plume. We conclude that the non-detection of the plumes in X-rays in
the {\it Chandra} observations is evidence that the bases of WAT
plumes are physically different from the bases of FRI type I jets; WAT
plumes are probably more similar to the large-scale jets of FRIs,
which typically do not show either strong deceleration or X-ray
emission.

It is not clear why we do not see X-ray emission associated with the
assumed rapid deceleration of the jet at the base of the plume, by
analogy with the X-ray emission detected from the hotspots of many
low-luminosity FRII radio galaxies, thought to be synchrotron in
origin (Hardcastle \etal\ 2004 and references therein). Features as
faint as the brightest knot at the end of the N jet (NJ2 in the
notation of Hardcastle \& Sakelliou: see also Fig. \ref{pbase}) have
been detected in X-ray observations of FRII hotspots. However, as we
noted above, there is no particularly good candidate feature analogous
to a hotspot in the radio images of 3C\,465; the jet flow clearly
continues after NJ2. {\it Chandra} images of WATs that {\it do}
exhibit compact bright jet termination structures would be of
considerable interest.

\subsection{WAT termination}

We begin by noting that the conclusions of Jetha \etal\ (2005), based
on less sensitive observations, apply to 3C\,465 as well. There is no
evidence in the data either for discrete features associated with the
jet termination points or plume bases or for discontinuities in the
properties of the external medium at the distances of the jet-plume
transition, such as the shock structures in the models of Loken \etal\
(1995); in particular, our observations show that the apparent
temperature discontinuity seen in the inner regions of the sources
studied by Jetha \etal\ was probably an unresolved temperature
gradient. Models that require a discontinuity in the external medium
are thus ruled out by observation. The jet termination is on a similar
scale to the transition between the central, cool X-ray core and the
hotter cluster medium, again as observed by Jetha \etal, but as yet we
have no clear idea of why this should be the case. As Jetha \etal\
suggest, numerical modelling of realistic (fast, low-density) jets and
plumes in realistic cluster environments may be the way forward here.
It is also interesting to note that the transition occurs close to the
point where the jet internal (minimum) pressure starts to exceed the
external thermal pressure (Section \ref{cluster} and Fig.\
\ref{pprofile}). X-ray and radio data of similar quality for a larger
number of sources would be required to say whether this is more than a
coincidence.

\subsection{Pressures and particle content}
\label{pcontent}
Our detailed X-ray and radio observations have given a very clear
indication that the large-scale plumes of 3C\,465 behave similarly to
the lobes or plumes of normal twin-jet FRI sources: the minimum
pressures in the lobes are about an order of magnitude below the
external thermal pressures. This implies some sort of departure from
the minimum-energy condition, either a departure from equipartition
between the radiating electrons and magnetic fields, a population of
non-radiating relativistic particles, or a contribution to pressure
from particles that do not participate in equipartition, such as
thermal protons. As usual (e.g. Hardcastle \etal\ 1998, Croston \etal\
2003) we cannot tell which of these is the case. In the inner parts of
normal FRI jets the jet deceleration is thought to be by entrainment
of external material, which has suggested that the entrained material
might supply the missing pressure if it could be heated efficiently so
as to avoid thermal X-rays from the lobes or plumes (see e.g. Croston
\etal\ 2003). In WAT plumes, the jets are not expected to have
entrained significant amounts of external material, since they do not
show the deceleration-related structures of FRI jets, and so this
explanation seems less likely (see also below, Section \ref{bending}).
On the other hand, if we assume that equipartition does not hold, it
is not clear why such a large departure from equipartition should be
required in WAT plumes but not in FRII lobes, where inverse-Compton
evidence (e.g. Hardcastle \etal\ 2002, Belsole \etal\ 2004, Croston
\etal\ 2004) suggests that equipartition is roughly maintained and
that the electrons and magnetic fields alone are sufficient to provide
pressure balance with the external medium. The departure from
equipartition would correspond to around a factor 5 difference between
the actual magnetic field strength and the equipartition value; this
is considerably larger than is typically seen in the X-ray-detected
lobes of FRII radio galaxies (Croston \etal\ 2005). In addition, a
departure from equipartition sufficient to make the internal pressure
equal to the external pressure in the direction of electron domination
would give rise to strong inverse-Compton emission from the plumes, at
the level of a few thousand {\it XMM} counts, which is not observed --
the deficits in X-ray count rate seen at the bases of the plumes seem
to rule out this model, so that the lobes would have to be
magnetically dominated if electrons and magnetic field alone were to
supply the required pressure. The third possibility is that the plumes
contain an energetically dominant population of relativistic protons:
this cannot be ruled out directly by the data, but again arguments
based on inverse-Compton detections suggest that it is not the case in
the hotspots (Hardcastle \etal\ 2004) or lobes (Croston \etal\ 2005)
of FRII sources. We return to the particle content of the plumes in
the next section.

\subsection{Bending}
\label{bending}

It is instructive to revisit the arguments of E84 that
led to the conclusion that the bending of 3C\,465 in particular, and
of WATs in general, was difficult to explain in any
model. They assumed that the kinetic energy of the plume must supply the
overall observed radio luminosity: that is, the bulk flow is cold and
there is {\it in situ} particle acceleration at all points in the
plume. This leads to a condition on the density and speed of a
non-relativistic plume:
\begin{equation}
L_{\rm rad} = \epsilon\pi r^2 \rho v_p^3/2
\label{lum}
\end{equation}
(their eq. 2) where $\epsilon$ is a factor giving the efficiency of
conversion between kinetic energy and energy of relativistic
electrons, $r$ is the plume radius, $\rho$ is its density and $v_p$ is
its speed. They assumed {\it in situ} particle acceleration on the
basis of estimates of the synchrotron loss timescale at 5 GHz, and so
we note to begin with that it may not be necessary to satisfy this
condition in its original form. The synchrotron age of material at the
end of the plumes is $\sim 1 \times 10^8$ years, based on 1.4 GHz and
330 MHz data taken from the VLA archives, if we assume a Jaffe \&
Perola (1973) aged synchrotron spectrum, an injection index of 2.0 and
a loss magnetic field strength around the minimum-energy value of 0.4
nT (taking into account CMB losses), and consequently material can be
transported out from the jets or plume bases without {\it in situ}
reacceleration so long as $v_p \ga 0.008c$, or 2500 km s$^{-1}$.

E84 used the lack of observed depolarization to place a limit ($n \la
80$ m$^{-3}$) on the density of thermal material (effectively,
protons) in the tails. This method of setting a density limit has
fallen out of favour in recent years, partly because it does not
provide a strong upper limit (Laing 1984) and partly because it seems
unlikely that the limits it does provide are close to the true values,
which may be much lower. It may thus not be appropriate to use a
proton density around $10$--$100$ m$^{-3}$ as a `typical' value in the
plumes. Observationally, there are few other constraints on the
internal density available: as discussed above, we know from the X-ray
deficits seen in both ordinary FRIs and in cluster-centre objects
(e.g. B\^\i rzan \etal\ 2004) that the missing pressure in these
sources is {\it not} supplied by thermal protons at the temperature of
the external medium, but this only really gives the constraint that
the internal density is likely to be much lower than the external one.
A true limit comes from considering the minimum possible internal
density. If we assume that the WAT plumes are in pressure balance and
that the internal pressure is supplied by relativistic particles and
magnetic field, presumably with some departure from the minimum energy
condition, the plume material remains very light compared to the
external medium: we require $nkT = U/3$, where $U = m_e c^2 \int
\gamma N(\gamma) {\rm d}\gamma + B^2/(2\mu_0)$, and the effective
density of the internal material is $U/c^2 = 3nkT/c^2$, while the
external density is $n\mu m_p$, where $\mu$ is the mass per particle
in atomic mass units (about 0.6). The density contrast is thus $\mu
m_p c^2/3kT$, or about $4 \times 10^4$ for the temperature of the
3C\,465 cluster. The internal density (in terms of an equivalent
proton density) in this model, which represents a minimum possible
internal density if we think that the plumes are not underpressured,
would be of the order of 0.04 m$^{-3}$, much lower than the `typical'
values used by E84. Such a light plume is comparatively easy --
perhaps even too easy -- to bend by ram pressure; if the flow speed up
the plume, $v_p$, were comparable to the local external sound speed
($c_s \sim 1300$ km s$^{-1}$), bulk motions with speeds of $\sim
0.005c_{\rm s}$ would have a significant effect, while if $v_p$ were
comparable to the speed estimated above from spectral ageing
arguments, we would require external bulk motions with speeds around
$0.01c_s$. If instead we adopt the luminosity condition (eq.
\ref{lum}) to estimate the flow speed, retaining this low internal
density, then we would obtain a bulk speed around $0.12c$ [for a
radiative efficiency of 0.1, which may be generous, in view of the jet
luminosity inferred for the weaker source 3C\,31 by Laing \& Bridle
(2002)] and, because of the low internal density, bulk motions with
speeds around $0.15c_s$ would still be sufficient to bend the plume:
as E84 comment, low jet densities are possible, but require high
speeds if the luminosity condition is to be satisfied.

If, on the other hand, the pressure deficit at minimum energy is
supplied by hot, entrained, internal thermal material, as is possible
for FRI lobes, then we know only that the density contrast is $T_{\rm
int}/T$ (since the thermal protons then dominate the density of the
flow) and in that case external bulk motions at the sound speed can
bend the plume significantly if $v_p/c_s \la \sqrt{T_{\rm int}/T}$,
which for moderate heating of the internal protons requires plume
speeds that are at most mildly supersonic: for example, if the plume
speed is 2500 km s$^{-1}$, then the internal protons must be heated to
around 20 keV (which would also give a density for internal protons
that was close to the `limit' on internal density inferred by E84).
Bending by subsonic motions requires a correspondingly higher
temperature, increasing as $(v/c_s)^{-2}$, but this is not impossible,
as we have little information on the temperature of the internal
protons. In the case of a proton-dominated plume the luminosity
condition, eq.\ \ref{lum}, requires $v_p \propto T_{\rm int}^{1/3}$,
and it can be shown that very high internal temperatures ($>3 \times
10^{10}$ K), correspondingly low densities, and high plume flow speeds
($>0.08c$) are required for the plume to be bent by sub-sonic bulk
motions in the external medium. To summarize, the plumes can be bent
by bulk motions in the external medium if they are light, but they
will only also meet the luminosity condition of E84 if they are also
fast. If we drop the assumption of {\it in situ} acceleration (and
instead assume passive transport of radiating electrons) then the
plumes need only be mildly supersonic, but must still be light.

This conclusion revives the possibility that the plumes are in fact
bent by the motion of the galaxy with respect to the host cluster. E84
suggest that the systematic motion of NGC 7720 could be of order
100--300 km s$^{-1}$, or $(0.08$--$0.23)c_s$. Their estimate does not
necessarily rule out higher galaxy speeds, because they base it on the
radial velocity of the galaxy with respect to the cluster mean (which
does not tell us about motions in the plane of the sky) and on studies
of the radial velocities of cDs in general (which may not be
representative of the systems hosting WATs). The possible detection of
an X-ray wake behind 3C\,465 (Section \ref{wake}) may indicate a
higher speed. Nevertheless, it is interesting that the speed estimate
of E84 is within the range estimated above for plume bending either if
the plume's internal pressure is supplied entirely by relativistic
particles and the plume is fast, or if a slower plume's internal
pressure is supplied by very hot thermal material. Is this picture
consistent with the fact that the inner jets are not bent? This
question is more difficult to answer because we do not have a reliable
estimate for the inner jet speed or density, and we cannot assume
pressure balance in the jet. If we assume $v_j \approx 0.5c$ (from the
beaming analysis) and take the minimum equivalent internal density to
be given by $\rho_j = 3p_{\rm min}/c^2$, using the minimum pressure
plotted in Fig.\ \ref{pprofile}, then we can use Euler's equation in
the form
\begin{equation}
{\rho_j v_j^2\over R} = {\rho_{\rm ext} v_g^2 \over h_j}
\end{equation}
where $R$ is the radius of curvature of the jet, $v_g$ is the galaxy
speed, and $h_j$ is the width of the jet. (The non-relativistic
approximation is not badly wrong in this case.) From the radio maps we can
see that the radio jet bends, if it bends at all, by less than one jet
width over its observed length $l$.
Taking $l=52$ arcsec (which includes the effects of projection with
$\theta = 50^\circ$), $h_j = 2$ arcsec, and an external density
comparable to that at the transition between the hot and cool
components in the X-ray (1500 protons m$^{-3}$) we find that speeds of
120 km s$^{-1}$ could produce a bend of one jet width. As this is very
consistent with the possible values for systematic motion of the
galaxy, and as our estimate of $\rho_j$ is a lower limit, we are
satisfied that an explanation for the plume bending in terms of the
systematic motion of the galaxy through the cluster is not
inconsistent with the observations of essentially straight inner jets,
provided that the jets are fast. We emphasise, however, that this
model is only viable if the plume is light. If there is substantial
entrained thermal material, then the analysis above suggests that
sub-sonic motion of the galaxy through the cluster material is too
slow to have a significant effect on the plumes. It is interesting in
this context that there is a population of WAT sources with bent inner
jets (exemplified by 0647+693, Hardcastle \& Sakelliou 2004): if the
jets in these sources {\it are} bent by ram pressure due to the host
galaxy's motion, then the model we have discussed suggests either that
the jets in these sources are weaker (transport less energy) or
slower, or that the systematic motions of the host galaxies are
larger. We will investigate this possibility in more detail elsewhere.

\label{fri-wat}
These results, combined with those of the previous subsection, mean
that there are some interesting possible differences between WATs and
classical FRI sources. On the one hand, we have shown (Section
\ref{pcontent}) that the large-scale plumes of 3C\,465 are very
similar to the extended structure of other FRI sources in terms of the
difference between the minimum internal pressure and the external
pressure: in FRIs one of the best explanations for this observation is
that the pressure deficit is supplied by heated, entrained material.
On the other hand, as we have seen, bending of the 3C\,465 plumes by
galactic motions puts limits on the amount of entrained material that
is present: for example, if we adopt a plume speed of 2500 km
s$^{-1}$, then significant bending by bulk speeds of $\sim 100$ km
s$^{-1}$ require thermal densities several orders of magnitude below
the external density (with correspondingly high temperatures if we
want this material to supply the missing pressure), while satisfying
the E84 luminosity condition requires even lower densities. These
densities are considerably lower than the densities estimated by O'Dea
(1985) from the bending of narrow-angle tail (NAT) sources. Although
it would be interesting to revisit the calculations of O'Dea with
modern estimates of the external density, and to compare them to
estimates of the proton density required to provide pressure in the
lobes, this may be an indication that type II jets such as those in
3C\,465 are less efficient at entraining thermal material than the
type I jets of NATs.

\section{Conclusions}

Our principal results can be summarized as follows:

\begin{enumerate}
\item The detection of X-ray emission from the inner part of the jet
  of 3C\,465 adds significantly to the evidence that fast (at least
  mildly relativistic), efficient, FRII-like (`type 2') jets can be
  sources of X-ray synchrotron radiation.
\item The non-detection of X-ray emission from the jet termination
  point and the bases of the plumes rules out a model in which the
  plume bases resemble the bright bases of FRI radio jets, which are
  often X-ray synchrotron sources: this is consistent with our
  favoured model for the WAT jet-plume transition, in which the
  deceleration of the fast inner jets takes place rapidly when the jet
  interacts with the external medium at the edge of the plume, rather
  than slowly by entrainment as in FRI jets. However, it is
  interesting that there is no detection of synchrotron X-rays from
  any of the plausible candidates for the jet termination `hotspot' in
  3C\,465, unlike the situation in the (supposedly physically similar)
  hotspots in low-power FRII sources. More X-ray observations of WATs with
  bright jet termination hotspots are required.
\item We have made detailed measurements of the temperature and
  density structure of 3C\,465's host cluster, A2634, and in
  particular have resolved a central dense, cool thermal component
  with a relatively short cooling time, which is similar to the
  components seen in other WATs and to the central components of FRI
  sources in groups. There is no evidence for any discontinuity in the
  properties of the external medium, and so we conclude, following
  Jetha \etal\ (2005), that models of WAT formation that require such
  discontinuities are ruled out. The most likely scenario based on our
  observations and those of Jetha \etal\ is that the location of the
  jet-plume transition is related to the emergence of the jet from the
  small-scale cool component associated with the host galaxy, but the
  physical mechanism that sets the location of the plume base remains unclear.
\item Our observations show that the minimum pressures of the plumes
  fall below the pressures estimated for the external thermal
  material: this implies either some departure from the minimum-energy
  (equipartition) condition or an additional contribution to pressure
  from low-density, hot thermal material. The plumes of WATs are
  thus similar to the large-scale components of more normal FRI
  sources.
\item Revisiting the arguments related to WAT plume bending, we find
  that the plumes can be bent by sub-sonic bulk motions of the ICM, or
  by plausible motion of the host galaxy through the ICM, provided
  that they are extremely light with respect to the external medium
  (as would be the case if their internal pressures were dominated by
  relativistic particles and/or magnetic field).
\end{enumerate}

\section*{Acknowledgments}

We thank Dan Evans for discussions of the X-ray core and helpful
comments on the paper, Nazirah Jetha for discussions of WAT
properties, and an anonymous referee for helpful comments. MJH thanks
the Royal Society for a research fellowship. The National Radio
Astronomy Observatory is a facility of the National Science Foundation
operated under cooperative agreement by Associated Universities, Inc.
This work is based on observations obtained with {\it XMM-Newton}, an
ESA science mission with instruments and contributions directly funded
by ESA Member States and NASA.

\end{document}